\documentclass[12pt,preprintnumbers,superscriptaddress,amsmath,amssymb,nofootinbib]{revtex4}

\usepackage{graphicx}
\usepackage{dcolumn}
\usepackage{bm}
\usepackage{amssymb}
\usepackage{amsmath}
\usepackage{epsfig}    
\usepackage{color}
\usepackage{slashed}

\def\be{\begin{equation}}
\def\ee{\end{equation}}
\newcommand{\bea}{\begin{eqnarray}}
\newcommand{\eea}{\end{eqnarray}}

\numberwithin{equation}{section}

\begin{document} 
\title{Higgs Phenomenology in the Minimal $SU(3)_L\times U(1)_X$ Model}

\author{Hiroshi Okada}
\email{hokada@kias.re.kr}
\affiliation{School of Physics, KIAS, Seoul 130-722, Korea}
\affiliation{Physics Division, National Center for Theoretical Sciences, Hsinchu, Taiwan 300}

\author{Nobuchika Okada}
\email{okadan@ua.edu}
\affiliation{Department of Physics and Astronomy, University of Alabama, Tuscaloosa, AL35487, USA}

\author{Yuta Orikasa}
\email{orikasa@kias.re.kr}
\affiliation{School of Physics, KIAS, Seoul 130-722, Korea}
\affiliation{Department of Physics and Astronomy, Seoul National University, Seoul 151-742, Korea}

\author{Kei Yagyu}
\email{K.Yagyu@soton.ac.uk}
\affiliation{
School of Physics and Astronomy, University of Southampton, Southampton, SO17 1BJ, United Kingdom}

\begin{abstract}

We investigate the phenomenology of a model based on the $SU(3)_c\times SU(3)_L\times U(1)_X$ gauge theory, the so-called 331 model. 
In particular, we focus on the Higgs sector of the model which is composed of three $SU(3)_L$ triplet Higgs fields, and this corresponds 
to the minimal form to realize phenomenologically acceptable scenario. 
After the spontaneous symmetry breaking $SU(3)_L\times U(1)_X\to SU(2)_L\times U(1)_Y$, 
our Higgs sector effectively becomes that with two $SU(2)_L$ doublet scalar fields, 
in which the first and the second generation quarks couple to the different Higgs doublet from 
that couples to the third generation quarks. 
This structure causes the flavour changing neutral current mediated by Higgs bosons at the tree level. 
By taking an alignment limit of the mass matrix for the CP-even Higgs bosons, which is naturally realized in the case
with the breaking scale of $SU(3)_L\times U(1)_X$ to be much larger than that of $SU(2)_L\times U(1)_Y$, 
we can avoid current constraints from flavour experiments such as the $B^0$-$\bar{B}^0$ mixing 
even for the Higgs bosons masses being ${\cal O}(100)$ GeV. 
In this allowed parameter space, we clarify that 
a characteristic deviation in quark Yukawa couplings of the standard model-like Higgs boson is predicted, 
which has a different pattern from that seen in two Higgs doublet models with a softly-broken $Z_2$ symmetry. 
We also find that the flavour violating decay modes of the extra Higgs boson, e.g., 
$H/A \to tc$ and $H^\pm \to ts$ can be dominant, and they yield the important signature to distinguish our model 
from the two Higgs doublet models. 

\end{abstract}
\maketitle

\section{Introduction}

The structure of the electroweak symmetry breaking $SU(2)_L\times U(1)_Y\to U(1)_{\text{em}}$ 
has been precisely tested by various collider experiments such as the LEP and SLC. 
Furthermore, the discovery of the Higgs boson at the CERN LHC supports the existence of an $SU(2)_L$ doublet scalar field which 
is required to realize the spontaneous breakdown of the electroweak symmetry in the minimal way. 
However, these facts do not necessarily mean that the $SU(2)_L\times U(1)_Y$ gauge symmetry describes 
the most fundamental theory.   
For example, models based on a larger gauge group containing the $SU(2)_L\times U(1)_Y$ subgroup can also explain the current experimental results. 

Among various possibilities for the extension of the electroweak symmetry, 
the choice of the $SU(3)_L\times U(1)_X$ group gives us an interesting consequence that 
the color triplet and the three generations for each type of fermions are related with each other due to the gauge anomaly cancellation~\cite{scheme1-1,scheme1-2}, 
while these two matters are irrelevant in the Standard Model (SM). 
So far, a variety of models based on $SU(3)_c\times SU(3)_L\times U(1)_X$, the so-called 331 models, 
have been discussed, where 
there are various ways to identify the electric charge $Q$ due to the rank two nature of the $SU(3)$ group and 
various embedding schemes of the SM fermions. 
We can classify these 331 models as listed in Table~\ref{xis}. 

In this paper, we study the phenomenology of a 331 model especially focusing on the Higgs sector. 
In our model, the Higgs sector is composed of three $SU(3)_L$ triplet scalar fields, 
which corresponds to the minimal choice to realize phenomenologically acceptable scenario\footnote{In fact, two $SU(3)_L$ triplets are enough to break the $SU(3)_L\times U(1)_X$ symmetry into $U(1)_{\text{em}}$, 
and such a model has been discussed in Ref.~\cite{two-triplets}. 
However in this configuration, the lightest up-type and down-type quarks become massless, so that 
it is difficult to reproduce the current data from flavour experiments.  }. 
After the breaking of $SU(3)_L\times U(1)_X$ into $SU(2)_L\times U(1)_Y$, 
our model can effectively be regarded as the two Higgs doublet model (THDM). 
The characteristic property of the Higgs sector 
is particularly seen in the structure of the Yukawa interactions, where 
the first and the second generation quarks couple to the different Higgs doublet from that couples to the third generation quarks. 
Although this inevitably causes the flavour changing neutral current (FCNC) mediated by Higgs bosons at the tree level, 
and it forces to set masses of the Higgs bosons to be typically ${\cal O}(10)$ TeV or larger.  
However, we find that 
we can avoid the current bound from flavour experiments even if we take masses of the Higgs bosons 
to be of the order of 100 GeV
by taking an alignment limit of the mass matrix of the CP-even Higgs bosons, which is naturally obtained in the limit where 
the breaking scale of $SU(3)_L\times U(1)_X\to SU(2)_L\times U(1)_Y$ to be infinity. 
 
In the allowed parameter regions, we first discuss the deviation in the SM-like Higgs boson couplings from the SM predictions. 
We clarify that our model predicts a characteristic pattern of the deviation in the quark Yukawa couplings which has a dependence on the quark flavour. 
This nature cannot be seen in THDMs with a softly-broken $Z_2$ symmetry. 
Next, we discuss the decay and production of the extra Higgs bosons at the LHC. 
We find that the flavour violating decay modes of the extra CP-even $H$ and CP-odd $A$ and singly-charged $H^\pm$ Higgs bosons can be dominant, e.g., 
$H/A \to tc$ and  $H^\pm \to ts$. 
Collider signatures from these decay modes provide us 
with an important tool to distinguish our model from the THDMs in addition to the deviation in the SM-like Higgs boson couplings. 

This paper is organized as follows. 
In Sec.~II, we define our minimal 331 model. We first present the particle content and the charge assignment. 
We then construct the kinetic Lagrangian for the scalar fields, the Higgs potential and the Yukawa Lagrangian. 
In Sec.~III, we take into account the current constraints on the parameter space from the LEP-II experiments and flavour data. 
In Sec.~IV, we discuss the Higgs phenomenology, i.e., the deviation in the SM-like Higgs boson couplings and the decay and production of the extra Higgs bosons. 
Conclusions are given in Sec.~V. 
In Appendices, we present the explicit analytic formulae for the Gauge-Gauge-Scalar type interaction terms (App.~A), 
the Higgs boson couplings to the SM fermions (App.~B), and the decay rate of the Higgs bosons (App.~C).

\begin{table}[t]
\begin{center}
{\renewcommand\arraystretch{1}
\begin{tabular}{c|c|c}\hline\hline
$\xi$ & Lepton triplet & Refs.\\\hline
$+1/\sqrt{3}$ & $(\mathbf{3}^*,-1/3)\sim (e^-,\nu,N)$ & \cite{scheme1-1,scheme1-2,scheme1-3}\\\hline
$+\sqrt{3}$ & $(\mathbf{3}^*,0)\sim (e^-,\nu,e^+)$ &  \cite{scheme3-1,scheme3-2}\\\hline
$-1/\sqrt{3}$ & $(\mathbf{3}^*,-2/3)\sim (e^-,\nu,E^{-})$ &  \cite{scheme5-1}\\\hline 
 $0$ & $(\mathbf{3}^*,-1/2)\sim  (e^-,\nu,E^{-1/2})$ & \cite{scheme0-1} \\\hline
$-\sqrt{3}$ & $(\mathbf{3},0)\sim (\nu,e^-,e^+)$ & \cite{scheme2-1,scheme2-2,scheme2-3,scheme2-4}\\\hline
$-1/\sqrt{3}$ & $(\mathbf{3},-1/3)\sim (\nu,e^-,N)$ & \cite{scheme4-0,scheme4-1,scheme4-2,scheme4-3}\\\hline
 \hline 
\end{tabular}}
\caption{Variations of the 331 model classified by $\xi$ and the embedding of lepton fields in the $(SU(3)_L,U(1)_X)$ multiplet, where 
$\xi$ determines the relation between the electric charge $Q$ and the $SU(3)$ Cartan generators given in Eqs.~(\ref{q}) and (\ref{cartan}). }
\label{xis}
\end{center}
\end{table}

\section{Model}

\subsection{Particle contents}

We discuss a model based on the gauge group $SU(3)_c\times SU(3)_L\times U(1)_X$. 
In this framework, there are several ways to identify the electric charge $Q$, because of the existence of two Cartan matrices of the $SU(3)$ group. 
Without loss of generality, $Q$ is defined as 
\begin{align}
Q = T_3 +\xi T_8 + X, \label{q}
\end{align}
where $X$ is the $U(1)_X$ charge, and $T_3$ and $T_8$ are the diagonal Gell-Mann matrices with the normalization of 
$\text{tr}(T^a T^b)=\delta^{ab}/2$:
\begin{align}
T_3 =\frac{1}{2} \text{diag}(1,-1,0),\quad T_8 = \frac{1}{2\sqrt{3}}\text{diag}(1,1,-2).  \label{cartan}
\end{align}
From Eq.~(\ref{q}), $Q$ is determined by specifying $\xi$ and $X$. 
When the SM left-handed lepton fields are embedded into the first and second components of a triplet or an anti-triplet representation of $SU(3)_L$, 
we have the following equations:
\begin{align}
\xi = \sqrt{3}(1 + 2X)~\text{if lepton triplet is $\mathbf{3}^*$}, \quad
\xi = -\sqrt{3}(1+ 2X)~\text{if lepton triplet is $\mathbf{3}$}. 
\end{align} 
In our model, we choose $\xi=1/\sqrt{3}$ and assign the left-handed leptons to be anti-triplet, which corresponds to 
the case listed in the first row of Table~\ref{xis}.  

\begin{table}[t]
\begin{center}
{\renewcommand\arraystretch{1}
\begin{tabular}{c|cccccc|ccc}\hline\hline
&\multicolumn{6}{c|}{Fermion Fields} & \multicolumn{3}{c}{Scalar Fields}\\\hline 
          & $Q_L^a$      & $Q_L^3$        &$(u_R^i,U_R)$      &$(d_R^i,D_R,S_R) $      & $L_L^i$      & $e_R^i$            &$\Phi_1$       &$\Phi_2$ & $\varphi$      \\\hline 
$SU(3)_c$ & $\mathbf{3}$ & $\mathbf{3}$   & $\mathbf{3}$&$\mathbf{3}$& $\mathbf{1}$ & $\mathbf{1}$  &  $\mathbf{1}$ & $\mathbf{1}$ & $\mathbf{1}$  \\\hline 
$SU(3)_L$ & $\mathbf{3}$ & $\mathbf{3^*}$& $\mathbf{1}$&$\mathbf{1}$ & $\mathbf{3}^*$& $\mathbf{1}$& $\mathbf{3}^*$ & $\mathbf{3}^*$& $\mathbf{3}^*$   \\\hline 
$U(1)_X $ & $0$          & $+1/3$        &$+2/3$       &$-1/3$       & $-1/3$  & $-1$              & $2/3$           & $-1/3$  & $-1/3$       \\\hline 
$Z_2 $    & $+$          & $+$           &$(+,-)$      &$(+,-,-)$    & $+$  & $+$              & $+$ & $+$  & $-$ 
\\\hline \hline 
\end{tabular}}
\caption{Particle content and its charge assignment under the $SU(3)_c\times SU(3)_L \times U(1)_X$ symmetry.
The indices $i$ and $a$ represent the flavour of fermions which run over 1-3 and 1-2, respectively. }
\label{particle}
\end{center}
\end{table}

The particle content is given in Table~\ref{particle}. 
In addition to the $SU(3)_c\times SU(3)_L\times U(1)_X$ gauge symmetry, we introduce a softly-broken discrete $Z_2$ symmetry which is required to 
avoid the mixing between SM quarks and exotic quarks. 

The fermion fields are parameterized as 
\begin{align}
Q_L^1 = \left(\begin{array}{c} 
u_L \\
d_L \\
D_L  
\end{array}\right),~
Q_L^2 = \left(\begin{array}{c} 
c_L \\
s_L \\
S_L  
\end{array}\right),~
Q_L^3 = \left(\begin{array}{c} 
b_L \\
t_L \\
U_L 
\end{array}\right),~
L_L^i = \left(\begin{array}{c} 
e_L^i \\
\nu_L^i \\
(N_R^i)^c 
\end{array}\right),~(i=1\text{-}3), \label{comp}
\end{align}
where $D_L$ and $S_L$ ($U_L$) are the left-handed exotic down (up)-type quarks with the electric charge of $-1/3$ ($2/3$). 
Similarly, $D_R$ and $S_R$ ($U_R$) are the right-handed exotic down (up)-type quarks. 

The scalar fields are parameterized by 
\begin{align}
\Phi_1 = \left(\begin{array}{c} 
\phi_1^0 \\
\phi_1^+ \\
\eta_1^+
\end{array}\right),~
\Phi_2 = \left(\begin{array}{c} 
\phi_2^- \\
\phi_2^0 \\
\eta_2^0
\end{array}\right), ~
\varphi = \left(\begin{array}{c} 
\eta_3^- \\
\eta_3^0 \\
\phi_3^0
\end{array}\right),
\end{align}
where the neutral components are expressed by
\begin{align}
&\phi_1^0 = \frac{h_1+ia_1+v_1}{\sqrt{2}},~
\phi_2^0 =\frac{h_2+ia_2+v_2}{\sqrt{2}},~
\phi_3^0 =\frac{h_3+ia_3^{}+u}{\sqrt{2}},  \label{component}\\
&\eta_2^0 =\frac{\eta_{R2}^{}+i\eta_{I2}^{}}{\sqrt{2}},~ 
\eta_3^0 =\frac{\eta_{R3}^{}+i\eta_{I3}^{}}{\sqrt{2}}. 
\end{align}
In Eq.~(\ref{component}), $v_1$, $v_2$ and $u$ are the vacuum expectation values (VEVs) for $\Phi_1$, $\Phi_2$ and $\varphi$, respectively. 
Under $v_1,v_2\ll u$, $v_1$ and $v_2$ determine the masses of the SM weak gauge bosons, while $u$ does 
the masses of extra gauge bosons. 
We will discuss the gauge boson masses in the next subsection. 
We note that the spontaneous symmetry breaking of $SU(3)_L\times U(1)_X$ occurs by the following steps:
\begin{align}
SU(3)_L\times U(1)_X \xrightarrow[u]{} SU(2)_L \times U(1)_Y \xrightarrow[v_1,v_2]{} U(1)_{\text{em}}, 
\end{align}
where the hypercharge $Y$ after the first step of the symmetry breaking is defined by $X+1/6$, $X-1/6$ and $X$ for 
the (originally) $SU(3)_L$ triplet, anti-triplet and singlet fields, respectively.

\subsection{Kinetic terms}

The kinetic term for the three $SU(3)_L$ triplet scalar fields are given by 
\begin{align}
\mathcal{L}_{\text{kin}} = |D_\mu \Phi_1|^2 + |D_\mu \Phi_2|^2 +  |D_\mu \varphi|^2,  
\end{align}
where $D_\mu$ is the covariant derivative. 
For a ${\bf 3}^*$ field with a $U(1)_X$ charge $X$, $D_\mu$ is given by 
\begin{align}
D_\mu   = \partial_\mu -ig(-T^{a*})A^a_\mu -ig_X^{}XB_\mu,~~ (a=1\text{-}8). 
\end{align}
The eight $SU(3)_L$ gauge bosons are expressed by the $3\times 3$ matrix form as 
\begin{align}
A_\mu \equiv A_\mu^aT^a = \frac{1}{2}\begin{pmatrix}
A_\mu^3 + \frac{1}{\sqrt{3}}A_\mu^8 & \sqrt{2}W^+_\mu & \sqrt{2}W^{\prime +}_\mu \\
\sqrt{2}W^-_\mu  & -A_\mu^3+\frac{1}{\sqrt{3}}A_\mu^8 & A^6_\mu - iA^7_\mu  \\
\sqrt{2}W^{\prime -}_\mu  & A^6_\mu + iA^7_\mu & -\frac{2}{\sqrt{3}}A_\mu^8
\end{pmatrix}, 
\end{align}
where
\begin{align}
W^\pm_\mu = \frac{1}{\sqrt{2}}(A^1_\mu \mp iA^2_\mu ),\quad 
W^{\prime \pm}_\mu = \frac{1}{\sqrt{2}}(A^4_\mu \mp iA^5_\mu ). \label{ch}
\end{align}
There are totally nine gauge bosons in this model, and they can be classified into 
2 pairs of massive charged gauge bosons expressed in Eq.~(\ref{ch}), and one (four) massless (massive) neutral gauge boson, where 
the massless gauge boson corresponds to the photon associated with the unbroken $U(1)_{\text{em}}$ symmetry. 

The squared masses of the charged gauge bosons $W^\pm$ and $W^{\prime \pm}$ are given by 
\begin{align}
m_W^2 =  \frac{g^2}{4}v^2, \quad m_{W'}^2 = \frac{g^2}{4}(v^2\cos^2\beta + u^2), \label{mwsq}
\end{align}
where $v=\sqrt{v_1^2+v_2^2}=(\sqrt{2}G_F)^{-1/2} \simeq 246$ GeV with $G_F$ being the Fermi constant, and 
$\tan\beta = v_2/v_1$. 
From the above formulae, we identify $W$ to be the SM $W$ boson with the mass of about 80 GeV, and $W'$ to be the extra charged gauge boson. 
In the following, we use the shorthand notation for an arbitrary angle $\theta$, i.e., $s_\theta=\sin \theta$, $c_\theta=\cos \theta$ and $t_\theta=\tan \theta$. 

For the neutral gauge bosons, it is convenient to define a basis where the photon state $A_\mu$ is separated from the other states as 
\begin{align} 
\begin{pmatrix}
A_3^\mu\\
A_8^\mu\\
B^\mu\\
A_6^\mu\\
A_7^\mu
\end{pmatrix} = 
\begin{pmatrix}
\frac{\sqrt{3}}{2}s_{331} & \frac{\sqrt{3}}{2}c_{331} & \frac{1}{2} & 0 & 0 \\
\frac{1}{2}s_{331} & \frac{1}{2}c_{331} & -\frac{\sqrt{3}}{2} & 0 & 0 \\
c_{331} & -s_{331} & 0 & 0 & 0 \\
0 & 0 & 0 & 1 & 0 \\
0 & 0 & 0 & 0 & 1 
\end{pmatrix}
\begin{pmatrix}
A^\mu\\
\tilde{Z}^\mu\\
\tilde{Z}^{\prime \mu}\\
Y_1^\mu\\
Y_2^\mu
\end{pmatrix},
  \label{base}
\end{align}
with $c_{331}=\cos\theta_{331}~\text{and}~ s_{331}=\sin\theta_{331}$ 
and $\tan\theta_{331} =2g_X^{}/(\sqrt{3}g)$. 
The mass matrix for the neutral gauge bosons in the basis shown in the right hand side of Eq.~(\ref{base}) is given as 
\begin{align}
{\cal M }_N^2 = 
\frac{g^2}{4}\begin{pmatrix}
0 & 0 & 0 & 0 & 0 \\
0& \frac{v^2(1+3c_\beta^2)+u^2}{3c^2_{331}}& \frac{v^2s_\beta^2-u^2}{\sqrt{3}c_{331}} & 0 &0 \\
0& \frac{v^2s_\beta^2-u^2}{\sqrt{3}c_{331}} & u^2+v^2s_\beta^2  & 0 & 0 \\
0&0&0 & u^2+v^2s_\beta^2 & 0 \\
0&0&0&0 & u^2+v^2s_\beta^2
\end{pmatrix}. \label{gauge_mass}
\end{align}
As we see Eqs.~(\ref{base}) and (\ref{gauge_mass}), the $\tilde{Z}_\mu$ and $\tilde{Z}_\mu'$ states are still not the mass eigenstates. 
We can define the mass eigenstates by introducing the mixing angle $\theta_Z$ as 
\begin{align}
\begin{pmatrix}
\tilde{Z} \\
\tilde{Z}' \\
\end{pmatrix}=
R(\theta_Z)
\begin{pmatrix}
Z \\
Z' \\
\end{pmatrix}, 
~~\text{with}~~
R(\theta) = \begin{pmatrix}
\cos\theta   & -\sin\theta \\
\sin\theta  &   \cos\theta
\end{pmatrix}. \label{rot}
\end{align}
The mixing angle $\theta_Z$ is given by 
\begin{align}
\tan2\theta_Z = \frac{2({\cal M}^2 _N)_{23}}{({\cal M}^2 _N)_{22}-({\cal M}^2 _N)_{33}} = \frac{2\sqrt{3}c_{331}(v^2s_\beta^2-u^2)}{4v^2c_\beta^2 +(u^2 + v^2s_\beta^2)(1-3c^2_{331})}. 
\label{thetaz}
\end{align}
Thus, the squared masses of all the neutral gauge bosons are expressed as 
\begin{align}
m_{Z,Z'}^2 &= \frac{1}{2}\left[
({\cal M}_N^2)_{22}+({\cal M}_N^2)_{33} \mp \sqrt{ \left[ ({\cal M}_N^2)_{22}-({\cal M}_N^2)_{33}\right]^2+4 ({\cal M}_N^2)_{23}^2}\right], \\
m_{Y_1}^2 &= m_{Y_2}^2 = \frac{g^2}{4}(u^2 + v^2s_\beta^2). 
\end{align}
Under $v^2/u^2\ll 1$, $m_Z^2$, $m_{Z'}^2$ and the mixing angle $\theta_Z$ are expanded by the series of $v^2/u^2$ as 
\begin{align}
m_{Z}^2 & = \frac{g^2}{1+3c_{331}^2 }v^2 + {\cal O}\left(\frac{v^4}{u^4} \right), \label{mzsq}\\
m_{Z'}^2 & = \frac{g^2u^2}{12c^2_{331}}\left[1+3c^2_{331} 
+\left(\frac{4}{1+3c_{331}^2}-3s_{331}^2 s_\beta^2\right)\frac{v^2}{u^2}\right] + {\cal O}\left(\frac{v^4}{u^4} \right), \\
\cos\theta_Z &= \sqrt{\frac{3}{1+3c_{331}^2}}-\frac{(1-3c_{331}^2)^2}{2(1+3c_{331}^2)^{5/2}}\frac{v^2}{u^2} + {\cal O}\left(\frac{v^4}{u^4} \right). \label{aaa}
\end{align}
In the limit of $u\to \infty$, the expression of $m_Z^2$ should be identical to the corresponding one in the SM, which allows us 
to identify the weak mixing angle $\theta_W$ in the SM as
\begin{align}
\cos\theta_W = \frac{1}{2}\sqrt{1+3c_{331}^2}. \label{bbb}
\end{align}
Using this expression, we reproduce $m_Z=gv/(2\cos\theta_W)$. 
The electroweak rho parameter is given to be unity in this limit:
\begin{align}
\rho_{\text{tree}}\equiv \frac{m_W^2}{m_Z^2 \cos^2\theta_W} = 1. 
\end{align}

In App.~A, we present the Gauge-Gauge-Scalar type interactions in the mass eigenstates of the Higgs bosons. 

\subsection{Higgs Potential}

The most general potential under the $SU(3)_L\times U(1)_X\times Z_2$ invariance is given by 
\begin{align}
V(\Phi_1,\Phi_2,\varphi) &= 
m_1^2|\Phi_1|^2 + m_2^2|\Phi_2|^2 + m_\varphi^2|\varphi|^2 
+ (m_{2\varphi}^2\Phi_2^\dagger \varphi - \mu \epsilon_{ABC}\Phi_1^A \Phi_2^B\varphi^C + \text{h.c.}) \notag\\
& + \frac{1}{2}\lambda |\varphi|^4 
  + \frac{1}{2}\lambda_1 |\Phi_1|^4
  + \frac{1}{2}\lambda_2 |\Phi_2|^4 
  + \lambda_3 |\Phi_1|^2|\Phi_2|^2
  + \lambda_4 |\Phi_1^\dagger \Phi_2|^2\notag\\
& + \sigma_1 |\Phi_1|^2|\varphi|^2
  + \sigma_2 |\Phi_1^\dagger \varphi|^2
  + \rho_1 |\Phi_2|^2|\varphi|^2
  + \rho_2 |\Phi_2^\dagger \varphi|^2
  + \frac{1}{2}[\rho_3 (\Phi_2^\dagger \varphi)^2+ \text{h.c.}], 
\end{align}
where the $\mu$ and $m_{2\varphi}^2$ terms are the soft breaking terms of the $Z_2$ symmetry. 
For the $\mu$ term, $A$, $B$ and $C$ $(=1$-3) are the indices of the $SU(3)_L$ fundamental space, and $\epsilon_{ABC}$ is the 
complete anti-symmetric tensor with $\epsilon_{123}=+1$. 
In the above potential, the $\mu$, $m_{2\varphi}^2$ and $\rho_3$ parameters are complex in general, while all the others are real. 
In the following, we take all the parameters to be real for simplicity. 

The tadpole terms for the neutral scalar states are given by 
\begin{align}
T_X \equiv \frac{\partial V}{\partial X}\Big|_0 , ~~\text{for}~~X=
h_\alpha^{},~
a_\alpha^{},~
\eta_{R2}^{},~\eta_{R3}^{},~
\eta_{I2}^{},~\eta_{I3}^{}~~ (\alpha=1,2,3), 
\end{align}
where
\begin{align}
T_{h_1}      &= \frac{v}{2}c_\beta(2m_{1}^2 +v^2c_\beta^2 \lambda_1 + v^2s_\beta^2\lambda_3 - u^2 \sigma_1+\sqrt{2}t_\beta\mu u), \\
T_{h_2}      &= \frac{v}{2}s_\beta(2m_{2}^2 +v^2s_\beta^2 \lambda_2 + v^2c_\beta^2\lambda_3 - u^2 \rho_1+\frac{\sqrt{2}}{t_\beta}\mu u), \\
T_{h_3}      &= \frac{u}{2}      (2m_{\varphi}^2 +u^2\lambda +v^2s_\beta^2\rho_1 + v^2\sigma_1 c_\beta^2 - \frac{\sqrt{2}\mu v^2}{u}c_\beta s_\beta), \\ 
T_{\eta_{R2}^{}} &= m_{2\varphi}^2 u, \quad
T_{\eta_{R3}}    = m_{2\varphi}^2v s_\beta, 
\end{align}
and all the other tadpoles are zero. 
By imposing the tadpole conditions $T_X=0$ for all $X$ under the assumption that all the VEVs $v_1$, $v_2$ and $u$ are non-zero, 
we can eliminate the parameters $m_{1}^2$, $m_2^2$, $m_\varphi^2$ and $m_{2\varphi}^2$, where 
the tadpole conditions from $T_{\eta_{R2}}$ and $T_{\eta_{R3}}$ give only one independent condition, i.e., $m_{2\varphi}^2=0$. 
Consequently, the Higgs potential is described by 14 independent parameters, i.e., 
5 (dimensionful parameters) plus 3 (VEVs) plus 10 (dimensionless parameters) minus 4 (independent tadpole conditions).

In the following, we discuss the masses of the Higgs bosons. 
In our model, there are totally $3\times 3\times 2=18$ scalar states, namely, four pairs of singly-charged states, five CP-odd states and five CP-even states. 
Among them, two pairs of singly-charged, three CP-odd states and one CP-even states correspond to the Nambu-Goldstone (NG) bosons which 
are absorbed into the longitudinal components of two pairs of charged gauge bosons ($W$ and $W'$) and four neutral gauge bosons ($Z$, $Z'$, $Y_1$ and $Y_2$). 
Therefore, we have two pairs of singly-charged Higgs bosons, one CP-odd  and three CP-even Higgs bosons as the physical states. 
It is important to mention here that 
the scalar states $\phi_{1,2,3}^0$ $(\phi_{1,2}^\pm)$ do not mix with $\eta_{2,3}^0$ $(\eta_3^\pm)$. 
This is because a kind of parity is remained after the $SU(3)_L$ breaking which is different from the $Z_2$ parity that is imposed to the Lagrangian.  
In addition, as we see in Sec.~\ref{yukawa}, 
these $\eta$ fields do not couple to the SM fermions. 
Therefore, the lightest neutral scalar component could be a candidate of dark matter. 
In this paper, we do not discuss the property of dark matter in detail, which is not the main target. 

We first discuss the masses for the 
parity even states under the residual symmetry. 
The mass eigenstates can be  defined by 
\begin{align}
\begin{pmatrix}
\phi_1^\pm \\
\phi_2^\pm
\end{pmatrix}
&=
R(-\beta)
\begin{pmatrix}
G^\pm \\
H^\pm
\end{pmatrix},~
\begin{pmatrix}
a_1 \\
a_2 \\
a_3 
\end{pmatrix}=R_{\text{odd}}
\begin{pmatrix}
G_{Z_1} \\
G_{Z_2} \\
A 
\end{pmatrix},~
\begin{pmatrix}
h_1\\
h_2\\
h_3
\end{pmatrix} = R
\begin{pmatrix}
H_1\\
H_2\\
H_3
\end{pmatrix}, \label{eigen}
\end{align}
where $R(\theta)$ is defined in Eq.~(\ref{rot}). 
$R_{\text{odd}}$ and $R$ are the orthogonal $3\times 3$ matrix, and the explicit form of the former one is given as 
\begin{align}
R_{\text{odd}} = \begin{pmatrix}
-\frac{n_1}{\sqrt{2}}(c_\beta + s_\gamma) & \frac{n_2}{\sqrt{2}}(c_\beta - s_\gamma) & \frac{s_\beta}{\sqrt{1 + s_\beta^2t_\gamma^2 }} \\
\frac{n_1}{\sqrt{2}}s_\beta  & -\frac{n_2}{\sqrt{2}}s_\beta & \frac{c_\beta}{\sqrt{1 + s_\beta^2t_\gamma^2}} \\
\frac{n_1}{\sqrt{2}}c_\gamma & \frac{n_2}{\sqrt{2}}c_\gamma & \frac{t_\gamma }{\sqrt{1 + s_\beta^2t_\gamma^2}} 
\end{pmatrix}, 
\end{align}
where $\tan\gamma = vc_\beta/u$, and $n_1$ and $n_2$ are the normalization factors:
\begin{align}
n_1 = (1+c_\beta s_\gamma)^{-1/2},\quad 
n_2 = (1-c_\beta s_\gamma)^{-1/2}. 
\end{align}
The rotation matrix of the CP-even states $R$ is generally expressed by three mixing angles. 
In Eq.~(\ref{eigen}), 
$G^\pm$ and $G^{'\pm}$ ($G_{Z_1}$ and $G_{Z_2}$) are the NG bosons which are absorbed into 
the longitudinal components of $W$ and $W'$, respectively (the linear combinations of $Z$ and $Z'$), while 
$H^\pm$, $A$ and $H_\alpha$ ($\alpha=1$-3) are the physical singly-charged, the CP-odd and the CP-even Higgs bosons, respectively. 
The squared masses of $H^\pm$ and $A$ are expressed by 
\begin{align}
m_{H^\pm}^2  = \frac{v^2}{2}\lambda_4 +M^2,~~
m_{A}^2     = M^2\left(1 + c_\beta^2\,t^2_\delta \right), 
\end{align}
where $\tan\delta =vs_\beta/u$, and $M^2 = \mu u/(\sqrt{2} s_\beta c_\beta)$.  
The squared masses of $H_\alpha$ are calculated 
from the $3\times 3$ mass matrix $M_H^2$ in the basis of $(h_1,h_2,h_3)$:
\begin{align}
M_{H}^2 &=  
\begin{pmatrix}
 v^2\lambda_1 c_\beta^2 + M^2s_\beta^2& (v^2\lambda_3-M^2) s_\beta c_\beta    & v(u\sigma_1c_\beta -\frac{\mu}{\sqrt{2}} s_\beta ) \\
(v^2\lambda_3 -M^2)s_\beta c_\beta & v^2\lambda_2 s_\beta^2 + M^2c_\beta^2  & v(u\rho_1s_\beta-\frac{\mu}{\sqrt{2}}c_\beta)    \\
v(u\sigma_1c_\beta -\frac{\mu}{\sqrt{2}} s_\beta )    & v(u\rho_1s_\beta-\frac{\mu}{\sqrt{2}}c_\beta)   &u^2\lambda  +\frac{v^2\mu}{u\sqrt{2}}s_\beta c_\beta\\
\end{pmatrix}. \label{mh}
\end{align}
Using $R$, the mass eigenvalues are expressed by 
\begin{align}
R^T\,M_{H}^2\,R &= \text{diag}(m_{H_1}^2\,m_{H_2}^2\,m_{H_3}^2). 
\end{align}

We here define an {\it alignment limit} of the mass matrix for the CP-even Higgs states $M_H^2$ as follows
\begin{align}
 u\sigma_1 - \frac{\mu}{\sqrt{2}}t_\beta  = 0, \quad
 u\rho_1 - \frac{\mu}{t_\beta\sqrt{2}}  = 0.  \label{align}
\end{align}
Under this alignment, the mass matrix $M_H^2$ becomes the block-diagonalized form with the $2\times 2$ and $1\times 1$ submatrices, 
and we obtain the following expression:
\begin{align}
R(\beta)^T (M_H^2)_{2\times 2} R(\beta) = M_H^{\prime 2},  \label{al1}
\end{align}
where 
\begin{align}
(M_H^{\prime 2})_{11} &= v^2(\lambda_1 c_\beta^4 + \lambda_2 s_\beta^4 +2\lambda_3 s_\beta^2 c_\beta^2), \\
(M_H^{\prime 2})_{22} &= v^2(\lambda_1 + \lambda_2  -2\lambda_3 )s_\beta^2 c_\beta^2+ M^2, \\
(M_H^{\prime 2})_{12} &= -v^2(\lambda_1 c_\beta^2 + \lambda_2 s_\beta^2 -\lambda_3 c_{2\beta})s_\beta c_\beta . 
\end{align}
We then obtain the analytic expressions for the mass eigenvalues and mixing angles as follows: 
\begin{align}
m_{H_1}^2 &= (M_H^{\prime 2})_{11}c_{\beta-\alpha}^2 + (M_H^{\prime 2})_{22}s_{\beta-\alpha}^2 - 2(M_H^{\prime 2})_{12}s_{\beta-\alpha}c_{\beta-\alpha}, \\
m_{H_2}^2 &= (M_H^{\prime 2})_{11}s_{\beta-\alpha}^2 + (M_H^{\prime 2})_{22}c_{\beta-\alpha}^2 + 2(M_H^{\prime 2})_{12}s_{\beta-\alpha}c_{\beta-\alpha}, \\
m_{H_3}^2 &= u^2\lambda + \frac{\mu}{\sqrt{2}u}v^2s_\beta c_\beta,   \label{al2}
\end{align}
where the mixing angle $\beta-\alpha$ is expressed by 
\begin{align}
\tan 2(\beta-\alpha) =  \frac{2(M_H^{\prime 2})_{12}}{(M_H^{\prime 2})_{22}-(M_H^{\prime 2})_{11}}. 
\end{align}
The rotation matrix $R$ is then expressed as 
\begin{align}
R = 
\begin{pmatrix}
c_\alpha & -s_\alpha & 0 \\
s_\alpha & c_\alpha & 0 \\
0 & 0 & 1
\end{pmatrix}. 
\end{align}
In the following, 
we use the two symbols for the two CP-even states, namely $(H_1,H_2)$ and $(H,h)$, and 
we identify $h$ as the Higgs boson discovered at the LHC with the mass of about 125 GeV, i.e., $m_h\simeq 125$ GeV. 
We note that the alignment limit is naturally realized by taking the limit of $v^2/u^2\to 0$. 
In this limit, the mass matrix given in Eq.~(\ref{mh}) can be expressed by the block diagonal form after taking 
an appropriate orthogonal transformation as 
\begin{align}
M_H^2 \to \begin{pmatrix}
(M_H^2)_{11} + {\cal O}(v^2) & (M_H^2)_{12} + {\cal O}(v^2) & 0 \\
(M_H^2)_{21} + {\cal O}(v^2) & (M_H^2)_{22} + {\cal O}(v^2) & 0 \\
0 & 0 & (M_H^2)_{33} + {\cal O}(v^2)
\end{pmatrix}, \label{align2}
\end{align}
where $(M_H^2)_{ij}$ are the matrix elements given in (\ref{mh}). 
Because the order $v^2$ corrections in the above expression can be absorbed by the reparametrization 
of the $\lambda$ parameters such as $\lambda_1$, $\lambda_2$, $\lambda_3$ and $\lambda$, we obtain
the essentially same result as given in Eqs.~(\ref{al1})-(\ref{al2}). 

Next, let us discuss the masses for the parity odd states. 
The mass eigenstates of them are defined as 
\begin{align}
\begin{pmatrix}
\eta_3^\pm \\
\eta_1^\pm
\end{pmatrix}
=
R(-\gamma)
\begin{pmatrix}
G^{'\pm} \\
\eta^\pm
\end{pmatrix},~
\begin{pmatrix}
\eta_{I3}^{} \\
\eta_{I2}^{} 
\end{pmatrix}=R(\delta)
\begin{pmatrix}
G_{Y_1}^0 \\
\eta_{I}^{}
\end{pmatrix}, ~
\begin{pmatrix}
\eta_{R3}\\
\eta_{R1}
\end{pmatrix} = R(-\delta)
\begin{pmatrix}
G_{Y_2}\\
\eta_R^{}
\end{pmatrix}, 
\end{align}
where
$G^{\prime \pm}$, $G_{Y_1}$ and $G_{Y_2}$ are the NG bosons absorbed into the 
longitudinal components of $W'$, $Y_1$ and $Y_2$. 
The squared masses are given by
\begin{align}
m_{\eta^\pm}^2  &= \frac{u^2}{2c_\gamma^2}\left(\sigma_2 + \frac{\sqrt{2}\mu}{u}t_\beta\right),~~
m_{\eta_{I,R}^{}}^2=  \frac{u^2}{2c_\delta^2}\left(\rho_2\mp \rho_3 + \frac{\sqrt{2}\mu }{ut_\beta}\right). 
\end{align}

It is important to mention here that in the limit of $v^2/u^2\to 0$, in which the alignment limit is naturally realized as explained in the above, 
$H^\pm$, $A$, $H$ and $h$ are remained in the low energy spectrum, but $H_3$, $\eta^\pm$, $\eta_{I}^{}$ and $\eta_{R}^{}$ are decoupled from the theory. 
As a result, our model effectively becomes the THDM. 


\subsection{Yukawa Lagrangian}\label{yukawa}

The Yukawa Lagrangians for the lepton (${\cal L}_L^Y$) and quark (${\cal L}_Q^Y$) sector are given by 
\begin{align}
-\mathcal{L}_L^Y &=
 \frac{1}{2} (Y_L)^{ij} (\overline{L_L^i})_A (L_L^{jc})_B  (\Phi_1^*)_C \epsilon^{ABC} 
+ (Y_e)^{ij}(\overline{L_L^i}) \Phi_1 e_R^j  + \text{h.c.}, \\
-\mathcal{L}_Q^Y &=
 (Y_{u1})^{ai}\overline{Q_L^a} \Phi_1^* u_R^i 
+(Y_{u2})^{i}\overline{Q_L^3} \Phi_2 u_R^i + Y_{U}\overline{Q_L^3} \varphi U_R + \text{h.c.}\notag\\
&+ (Y_{d1})^{i}\overline{Q_L^3}\Phi_1 d_R^i + (Y_{d2})^{ai}\overline{Q_L^a} \Phi_2^* d_R^i + (Y_D)^{am}\overline{Q_L^a} \varphi^* D_R^m + \text{h.c.}, 
\end{align}
where $D_R^{m=1}=D_R$ and $D_R^{m=2}=S_R$, and 
the $Y_L$ coupling is the anti-symmetric $3\times 3$ matrix. 
This term gives the mixing among the component fields of $L_L$, i.e., $\nu_L^{}$-$N_L^c$ (see Eq.~(\ref{comp})). 
Because of the anti-symmetric structure of $Y_L$, it is not sufficient to reproduce the current neutrino oscillation data. 
However, as discussed in Ref.~\cite{scheme1-3}, if we introduce additional $SU(3)_L$ singlet neutral fermions, one-loop induced Majorana neutrino 
masses appear, and then the neutrino data can be reproduced. 
In this paper, we do not discuss the neutrino sector, and we take $Y_L$ negligibly small. 

The mass matrices for the charged leptons $(\mathcal{M}_e)$, the up-type quarks ($\mathcal{M}_u$) and 
the down-type quarks ($\mathcal{M}_d$) are respectively given by the $3\times 3$, $4\times 4$ and $5\times 5$ form as  
\begin{align}
-{\cal L}^{\text{mass}} &= 
\overline{\vec{E}_L} {\cal M}_e \vec{E}_R
+\overline{\vec{U}_L} {\cal M}_u \vec{U}_R
+\overline{\vec{D}_L} {\cal M}_d \vec{D}_R 
+ \text{h.c.},
\end{align}
where $\vec{E}_{L,R}=(e,\mu,\tau)_{L,R}$, 
$\vec{U}_{L,R}=(u,c,t,U)_{L,R}$ and $\vec{D}_{L,R}=(d,s,b,D,S)_{L,R}$. 
The form of ${\cal M}_e$ is the same as in the SM, i.e., ${\cal M}_e = v_1Y_e/\sqrt{2}$. 
On the other hand, ${\cal M}_u$ and ${\cal M}_d$ take 
the block-diagonalized form due to the $Z_2$ symmetry, where 
the first $3\times 3$ part corresponds to the mass matrix for the SM quarks, and 
the latter part does to that for the exotic quarks ($1\times 1$ for up-type and $2\times 2$ for down-type exotic quarks). 
Namely, 
\begin{align}
{\cal M}_u & = \frac{\text{diag}(v_1,v_1,v_2,u)}{\sqrt{2}}
\begin{pmatrix}
Y_u^{\text{SM}} & 0 \\
0   & Y_U 
\end{pmatrix},~
{\cal M}_d  = \frac{\text{diag}(v_2,v_2,v_1,u,u)}{\sqrt{2}}
\begin{pmatrix}
Y_{d}^{\text{SM}} & 0 & 0\\
0& Y_{D}^{11} & Y_{D}^{12}\\
0& Y_{D}^{21} & Y_{D}^{22}
\end{pmatrix}, 
\end{align}
where
\begin{align}
Y_u^{\text{SM}} = 
\begin{pmatrix}
Y_{u1}^{11} & Y_{u1}^{12}  & Y_{u1}^{13} \\
Y_{u1}^{21} & Y_{u1}^{22}  & Y_{u1}^{23} \\
Y_{u2}^{1}  & Y_{u2}^{2}   & Y_{u2}^{3} 
\end{pmatrix}, ~
Y_d^{\text{SM}} = 
\begin{pmatrix}
 Y_{d2}^{11} &  Y_{d2}^{12} & Y_{d2}^{13} \\
 Y_{d2}^{21} &  Y_{d2}^{22} & Y_{d2}^{23} \\
 Y_{d1}^{1}  &  Y_{d1}^{2}  & Y_{d1}^{3}  
\end{pmatrix}. 
\end{align}

The interaction terms for the SM quarks (${\cal L}_Q^{\text{int}}$) and 
those for the SM leptons (${\cal L}_L^{\text{int}}$) with a Higgs boson are expressed in their mass eigenbasis as
\begin{align}
{\cal L}_Q^{\text{int}} 
&= \frac{1}{v}\sum_{\phi=H_1,H_2,H_3,A}\left[\overline{d_L^i}\, (\Gamma_d^\phi)^{ij}\,    d_R^{j}  
+ \overline{u_L^i}\, (\Gamma_u^\phi)^{ij}\,    u_R^{j}\right]\phi + \text{h.c.} \notag\\
&+\frac{\sqrt{2}}{v}\left[ \overline{u^i_L} \,(\Gamma_d^{H^\pm})^{ij} \,d_R^j   +
\overline{u_R^i}\,(\Gamma_u^{H^\pm\,\dagger})^{ij} \,d_L^j \right]H^+ + \text{h.c.} \label{yint}\\
{\cal L}_L^{\text{int}} &=\frac{m_{e^i}}{v}\overline{e_L^i}e_R^j\left[\sum_{\alpha=1,3}\frac{R_{1\alpha}}{c_\beta} H_\alpha +i\frac{t_\beta}{\sqrt{1+s_\beta^2 t_\gamma^2}}A\right]
+\frac{\sqrt{2}m_{e^i}}{v}\overline{\nu_L^i}\, e_R^j H^+ + \text{h.c.}, 
\end{align}
where $\Gamma_q^\phi$ and $\Gamma_q^{H^\pm}$ ($q=u,d$) are the $3\times 3$ form of the dimensionful couplings. 
All the analytic expressions of them are given in App.~B. 
It is important to mention here that the $\Gamma_q^\phi$ couplings generally contain non-zero off-diagonal elements, so that 
the tree level FCNCs appear via the Higgs boson mediations. 
We will see in Sec.~IV that by taking the alignment limit and $\sin(\beta-\alpha)=1$, 
$\Gamma_q^h$ become diagonal, and thus the tree level FCNCs mediated by $h$ disappear. 
On the other hand, $\Gamma_q^H$ and $\Gamma_q^A$
have non-zero off-diagonal elements even in this limit. 
As a result, $H$ and $A$ contribute to FCNC processes. 
We will discuss the constraint on the parameter space from neutral meson mixings such as $B^0$-$\bar{B}^0$ in Sec.~III-B.

\section{Constraints}

\begin{figure}[t]
\begin{center}
 \includegraphics[width=100mm]{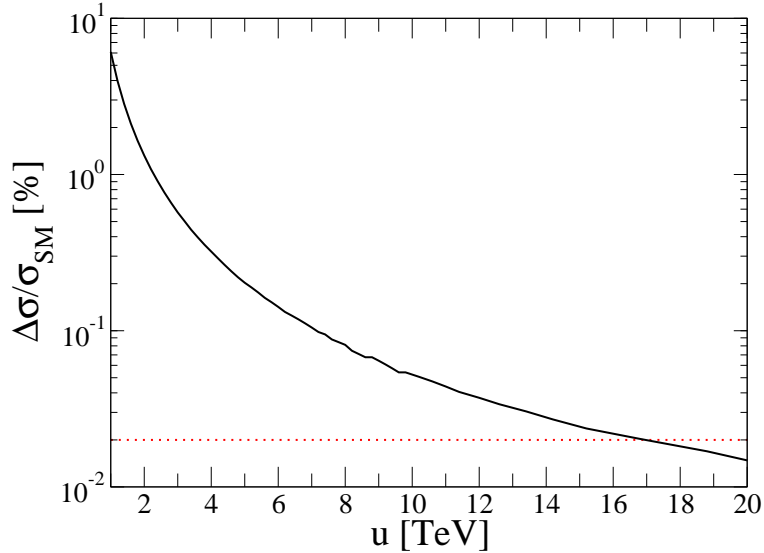}
   \caption{Deviation in the cross section of $e^+e^-\to \mu^+\mu^-$ process 
as a function of $u$ at the center of mass energy of 200 GeV.  
The horizontal dotted line shows the upper bound on the deviation at 95\% CL. 
 }
   \label{LEP2}
\end{center}
\end{figure}

In this section, we discuss constraints on the parameter space from experimental data. 
We first take into account the constraint from the LEP-II experiments, and then 
we consider that from flavour experiments. 

\subsection{LEP-II}

The  $e^+e^- \to f\bar{f}$ processes have been precisely measured at the LEP-II experiments by the 
center of mass energy of around 200 GeV, which 
derives a strong bound on the VEV $u$ describing the breaking scale of $SU(3)_L\times U(1)_X \to SU(2)_L\times U(1)_Y$. 
In Ref.~\cite{Abbiendi:2003dh}, the deviations in this cross section from the SM prediction are given at the center of mass energy to be between 189 GeV and 209 GeV. 
Among the various final states, the $\mu^+ \mu^-$ channel is most accurately measured whose one standard deviation has been given to be 0.01\%. 

In our model, the cross section of the $e^+e^- \to f\bar{f}$ process can be deviated from the SM prediction by the following sources, 
(i) the deviation in the $Z$-$f$-$\bar{f}$ coupling, (ii) the contribution from the $Z'$ boson exchange, and (iii) 
the interference effects between the $Z$ and $\gamma$ contributions and the $Z'$ contribution. 
In order to calculate the cross section, we extract the $\bar{f}f V_\mu$ vertex ($V_\mu=A_\mu$, $Z_\mu$ and $Z_\mu'$, and $f$ being the SM fermion) as
\begin{align}
{\cal L}_{ffV} &= eQ_f\bar{f}\gamma^\mu f A_\mu 
+ g_Z^{} \bar{f}\gamma^\mu \left(I_f -Q_f\sin^2\theta_W \right) f Z_\mu \notag\\
&-g_Z^{}\tan\theta_Z \bar{f}\gamma^\mu \left(I_f - Q_f\sin^2\theta_W \right)f Z_\mu', \label{ffz}
\end{align}
where $I_f=+1/2\,(-1/2)$ for $f=u\,(d,e)$, and 
\begin{align}
e=\frac{\sqrt{3}}{2}gs_{331}=g\sin\theta_W, \quad 
g_Z^{}=\frac{2g}{\sqrt{3}c_{331}}\cos\theta_Z = \frac{2g}{\sqrt{4\cos^2\theta_W-1}}\cos\theta_Z.
\end{align}
We note that in the limit of $v^2/u^2 \to 0$, we reproduce the SM $\bar{f}$-$f$-$Z_\mu$ coupling, i.e., 
$g_Z^{} \to g/\cos\theta_Z$ by using Eqs.~(\ref{aaa}) and (\ref{bbb}). 
From Eq.~(\ref{ffz}), the deviation in the cross section depends on the angle $\theta_Z$ which is determined by $u$ and $\tan\beta$ 
as shown in Eq.~(\ref{thetaz}). 

In Fig.~\ref{LEP2}, we plot the prediction of the deviation in the cross section of $e^+e^-\to \mu^+\mu^-$ represented by 
$\Delta \sigma$ as a function of $u$. 
We define $\Delta \sigma$ as 
\begin{align}
\Delta \sigma \equiv \sigma_{\text{331 Model}}-\sigma_{\text{SM}}, 
\end{align}
where $\sigma_{\text{331 Model}}$ ($\sigma_{\text{SM}}$) is the cross section of $e^+e^-\to \mu^+\mu^-$ in our model (SM). 
The horizontal line represents the 95\% CL upper limit on the deviation for the cross section. 
Although the $\tan\beta$ dependence on $\Delta \sigma$ is negligibly small when $v^2/u^2\ll 1$, we take $\tan\beta=1$ in this plot.  
We use {\tt CalCHEP}~\cite{calchep} for the numerical evaluation of the cross section. 
By looking at the cross point of two curves, we obtain the lower limit of $u \gtrsim 17$ TeV at 95\% CL.

\subsection{FCNC}

As we mentioned in Sec.~II-D, there appear the flavour violating Yukawa couplings at the tree level. 
Therefore, we expect to get a severe constraint on parameters from data at flavour experiments. 

In this subsection, we calculate the contributions to the mixing in neutral mesons such as $K^0$-$\bar{K}^0$ via the neutral Higgs boson mediations. 
The relevant effective Hamiltonian ${\cal H}_{\text{eff}}$ to these processes is given by
\begin{align}
{\cal H}_{\text{eff}} &= \sum_{i,j}c_{ij}^{}{\cal O}_{ij},\quad (i,j)=(L,R), \label{effH1}
\end{align}
where $c_{ij}$ and ${\cal O}_{ij}$ are the Wilson coefficients and dimension 6 operators, respectively. 
For the case of the $K^0$ and $\bar{K}^0$ mixing as an example, these operators are expressed by 
\begin{align}
{\cal O}_{ij} = (\bar{d}^\alpha P_i s^\alpha)(\bar{d}^\beta P_j s^\beta),  \label{effH2}
\end{align}
where $\alpha$ and $\beta$ are the color indices, and $P_{L,R}$ are the left- and right-handed projection operator. 
The matrix element of ${\cal O}_{ij}$ for the $K^0$ and $\bar{K}^0$ state is given by \cite{Gabbiani} 
\begin{align}
&\langle K^0 |{\cal O}_{LL}| \bar{K}^0 \rangle =\langle K^0 |{\cal O}_{RR}| \bar{K}^0 \rangle = -\frac{5}{24}\left(\frac{m_K^{}}{m_s+m_d} \right)^2m_K^{}f_K^2, \label{effH3} \\
&\langle K^0 |{\cal O}_{LR}| \bar{K}^0 \rangle =\langle K^0 |{\cal O}_{RL}| \bar{K}^0 \rangle = \left[\frac{1}{24} + \frac{1}{4}\left(\frac{m_K^{}}{m_s+m_d}\right)^2 \right]m_K^{}f_K^2, 
\label{effH4}
\end{align}
where $m_d$, $m_s$ and $m_K^{}$ are the masses of the down quark, the strange quark and the $K$ meson, respectively, 
and $f_K$ is the decay constant of the $K$ meson. 
The $K^0$-$\bar{K}^0$ mixing parameter $\Delta m_K^{}$ is calculated by using the above parameters as:
\begin{align}
\Delta m_K &= 2\text{Re}\langle K^0 |{\cal H}_{\text{eff}}| \bar{K}^0 \rangle \notag\\
&= \left\{c_{LR}^{} \left[\frac{1}{6} + \left(\frac{m_K^{}}{m_s+m_d}\right)^2 \right] 
-\frac{5}{12}(c_{LL}^{} + c_{RR}^{})\left(\frac{m_K^{}}{m_s+m_d} \right)^2
\right\}m_K^{}f_K^2 . \label{effH5}
\end{align}
Similarly, we obtain the predictions for the other meson mixings, namely, 
the $B^0$-$\bar{B}^0$ mixing $\Delta m_B$ and the $D^0$-$\bar{D}^0$ mixing $\Delta m_D$ are respectively obtained by the replacement of 
$(m_K,f_K,\bar{m}_s )\to (m_B,f_B,\bar{m}_b)$ and $(m_K,f_K,\bar{m}_s )\to (m_D,f_D,\bar{m}_c)$. 

Let us express the coefficients $c_{ij}$ in terms of the Lagrangian parameters. 
These are expressed for the $K^0$-$\bar{K}^0$ mixing:
\begin{align}
c_{LL}^{} = \sum_{\phi=h,H,A}\frac{(\Gamma_d^{\phi *})_{21}^2}{m_{\phi}^2v^2}, ~
c_{RR}^{} = \sum_{\phi=h,H,A}\frac{(\Gamma_d^{\phi})_{12}^2}{m_{\phi}^2v^2},  ~
c_{LR}^{} = c_{RL}^{} = \sum_{\phi=h,H,A}\frac{(\Gamma_d^{\phi *})_{21}(\Gamma_d^{\phi})_{12}}{m_{\phi}^2v^2},  \label{cc1}
\end{align}
for the $B^0$-$\bar{B}^0$ mixing: 
\begin{align}
c_{LL}^{} = \sum_{\phi=h,H,A}\frac{(\Gamma_d^{\phi *})_{31}^2}{m_{\phi}^2v^2}, ~
c_{RR}^{} = \sum_{\phi=h,H,A}\frac{(\Gamma_d^{\phi})_{13}^2}{m_{\phi}^2v^2},  ~
c_{LR}^{} = c_{RL}^{} = \sum_{\phi=h,H,A}\frac{(\Gamma_d^{\phi *})_{31}(\Gamma_d^{\phi})_{13}}{m_{\phi}^2v^2}, \label{cc2}
\end{align}
and for the $D^0$-$\bar{D}^0$ mixing:
\begin{align}
c_{LL}^{} = \sum_{\phi=h,H,A}\frac{(\Gamma_u^{\phi *})_{12}^2}{m_{\phi}^2v^2}, ~
c_{RR}^{} = \sum_{\phi=h,H,A}\frac{(\Gamma_u^{\phi})_{21}^2}{m_{\phi}^2v^2},  ~
c_{LR}^{} = c_{RL}^{} = \sum_{\phi=h,H,A}\frac{(\Gamma_u^{\phi *})_{12}(\Gamma_d^{\phi})_{21}}{m_{\phi}^2v^2}.  \label{cc3}
\end{align}

In order to evaluate $\Delta m_K$, $\Delta m_B$ and $\Delta m_D$ numerically, we use the following
input values given in MeV as~\cite{PDG,HFAG}:
\begin{align}
&m_{K} = 497.611,~\Delta m_K = 3.484\times 10^{-12},~ f_K^{} = 156.3,~ \bar{m}_s(m_s) = 95, \notag\\
&m_{D}^{}=1864.84,~
\Delta m_D = 6.25\times 10^{-12},~ f_D^{} = 212.6,~ \bar{m}_c(m_c) = 1275, \notag\\
&m_{B} = 5279.61,~\Delta m_B = 3.356\times 10^{-10},~ f_B^{} = 190.5,~ \bar{m}_b(m_b) = 4180, \label{exp}
\end{align}
For the unitary matrices of the left-handed quarks $V_L^q$($q=u,d$) we use the following values, 
\begin{align}
V_L^u &=
\begin{pmatrix}
0.975 & -0.223 & 1.86\times 10^{-3} \\
0.222 & 0.974  & 0.0518 \\
-0.01340& -0.0501& 0.999
\end{pmatrix}
+i
\begin{pmatrix}
2.83\times 10^{-6} &1.24\times 10^{-5}&- 1.79\times 10^{-3}\\
- 1.03\times 10^{-4}& 2.35\times 10^{-5}& 0 \\
- 1.74\times 10^{-3}& 3.98\times 10^{-3}& 0
\end{pmatrix}, \notag\\
V_L^d&=
\begin{pmatrix}
1.00&2.56\times 10^{-3}&5.87\times 10^{-3}\\
-3.10\times 10^{-3}&0.996&0.0941\\
-5.61\times 10^{-3}&-0.0942&0.996
\end{pmatrix}
-i
\begin{pmatrix}
0&0& 4.11\times 10^{-3}\\
 3.87\times 10^{-4}& 9.91\times 10^{-7}&0\\
 4.09\times 10^{-3}& 1.05\times 10^{-5}&0
\end{pmatrix}, \label{qmatrix}
\end{align}
by which the experimental values of the elements of Cabibbo-Kobayashi-Maskawa (CKM) matrix~\cite{PDG} defined as $V_L^u (V_L^{d})^\dagger$ are reproduced.

\begin{figure}[t]
\begin{center}
\includegraphics[width=55mm]{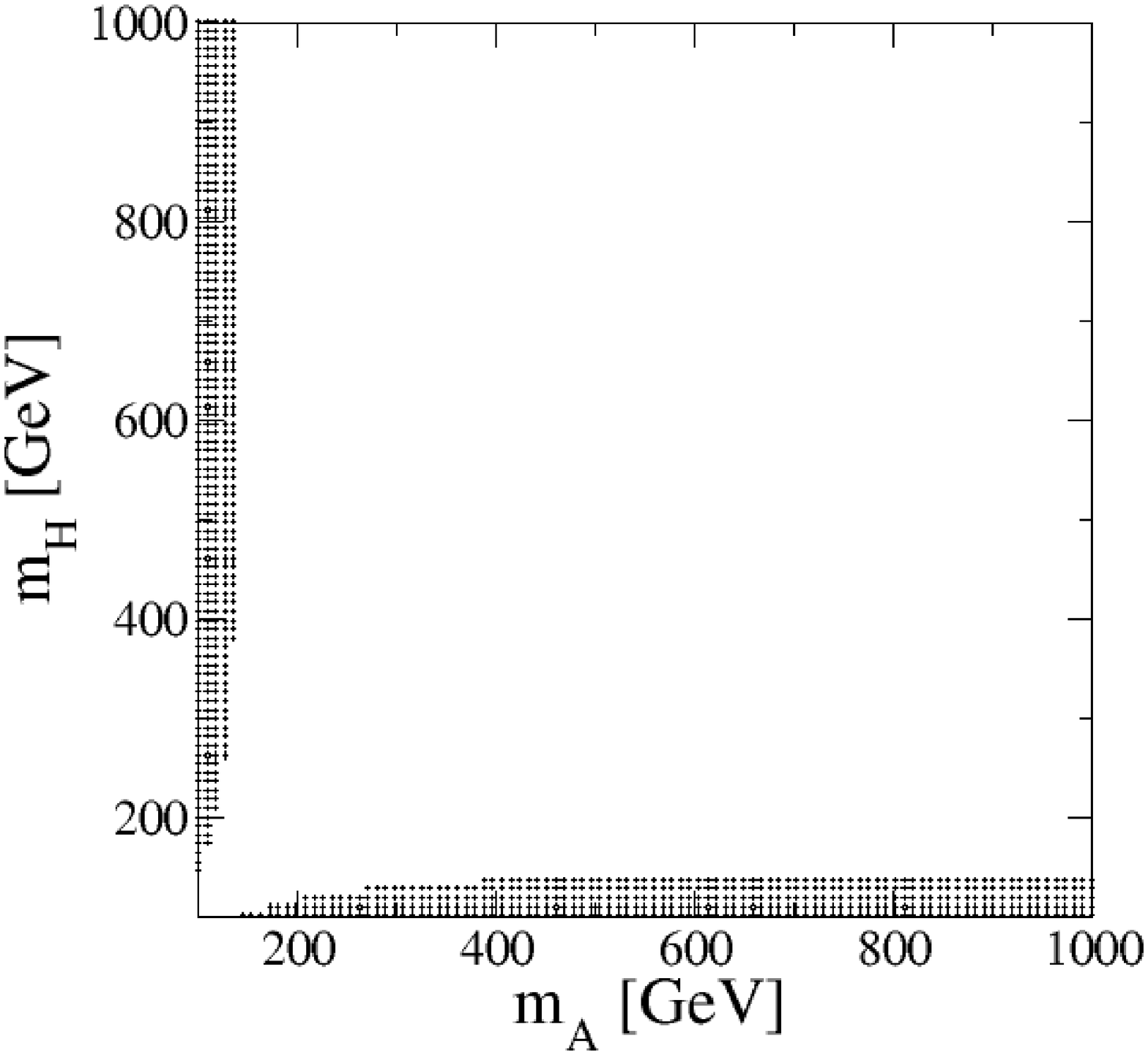}  \hspace{-5mm}
\includegraphics[width=55mm]{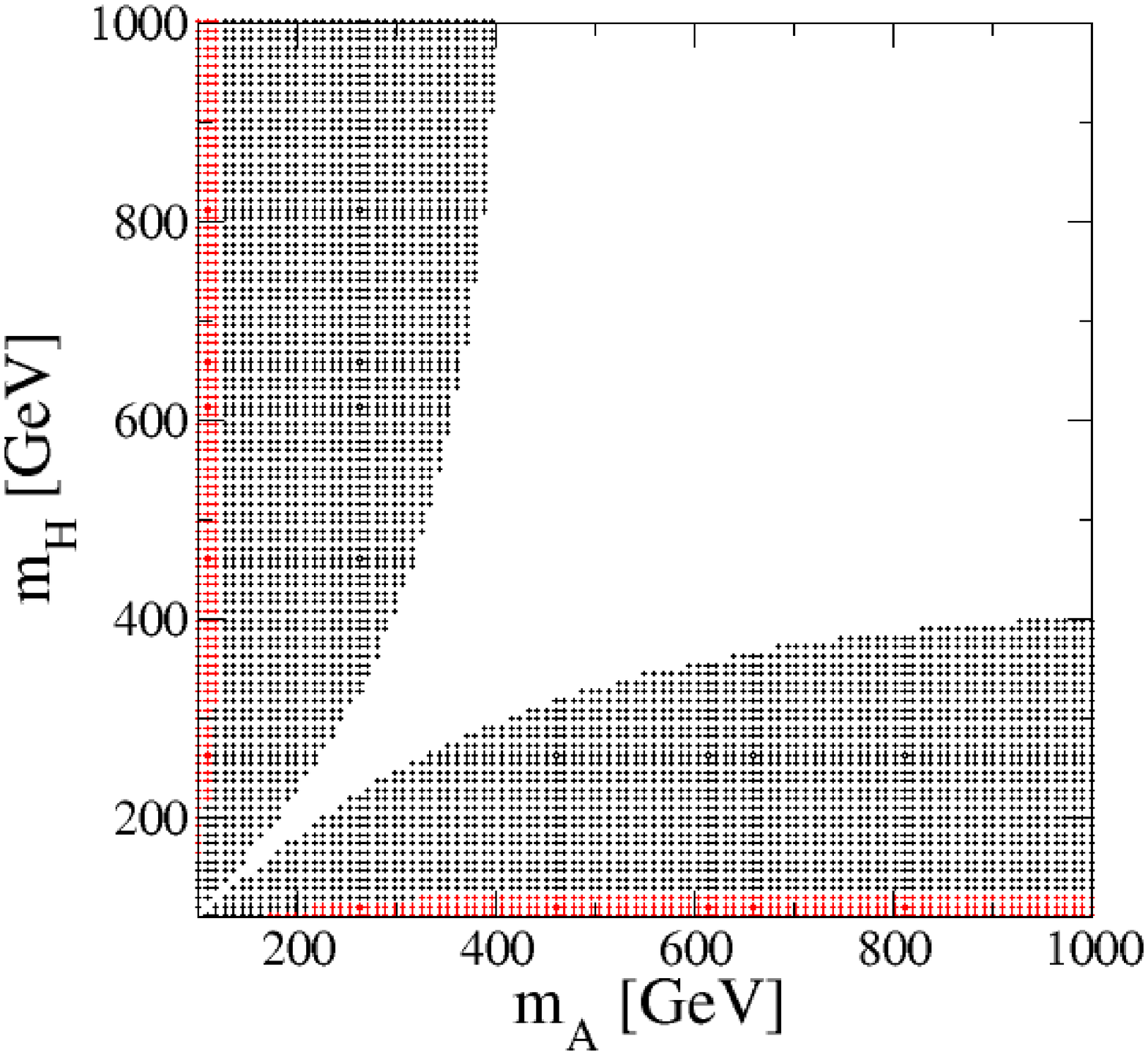} \hspace{-5mm}
\includegraphics[width=55mm]{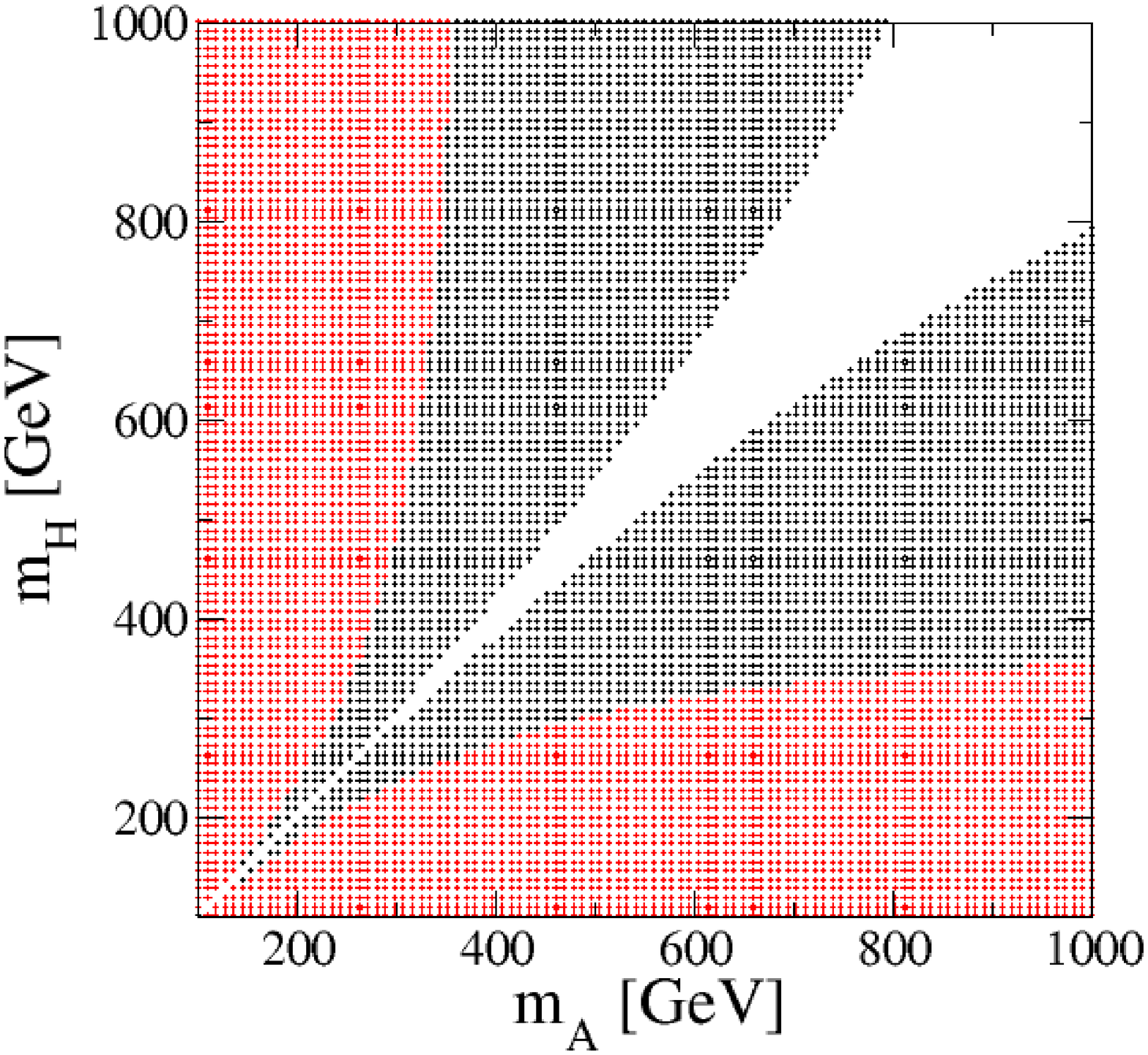}\\
\includegraphics[width=55mm]{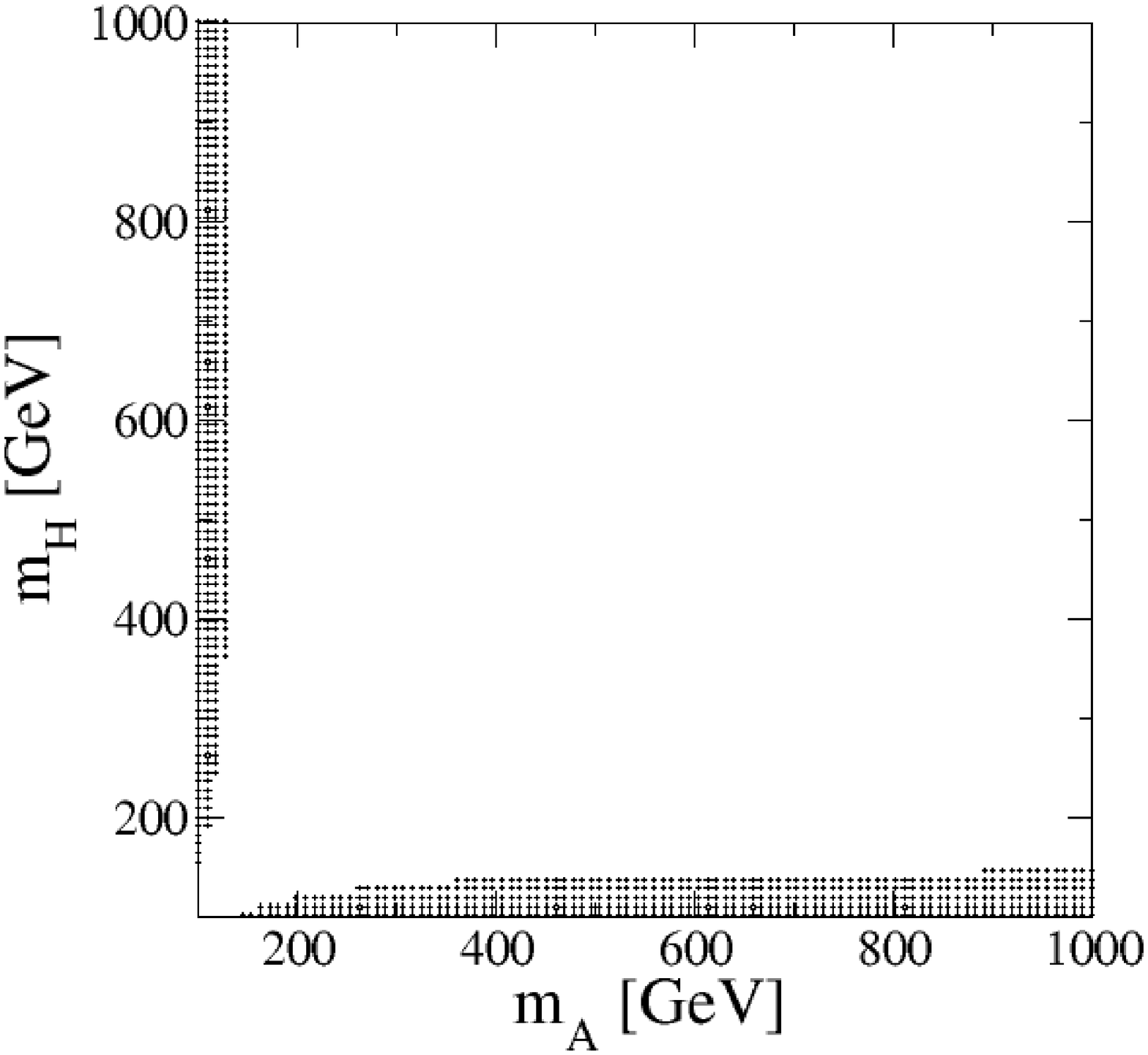}  \hspace{-5mm}
\includegraphics[width=55mm]{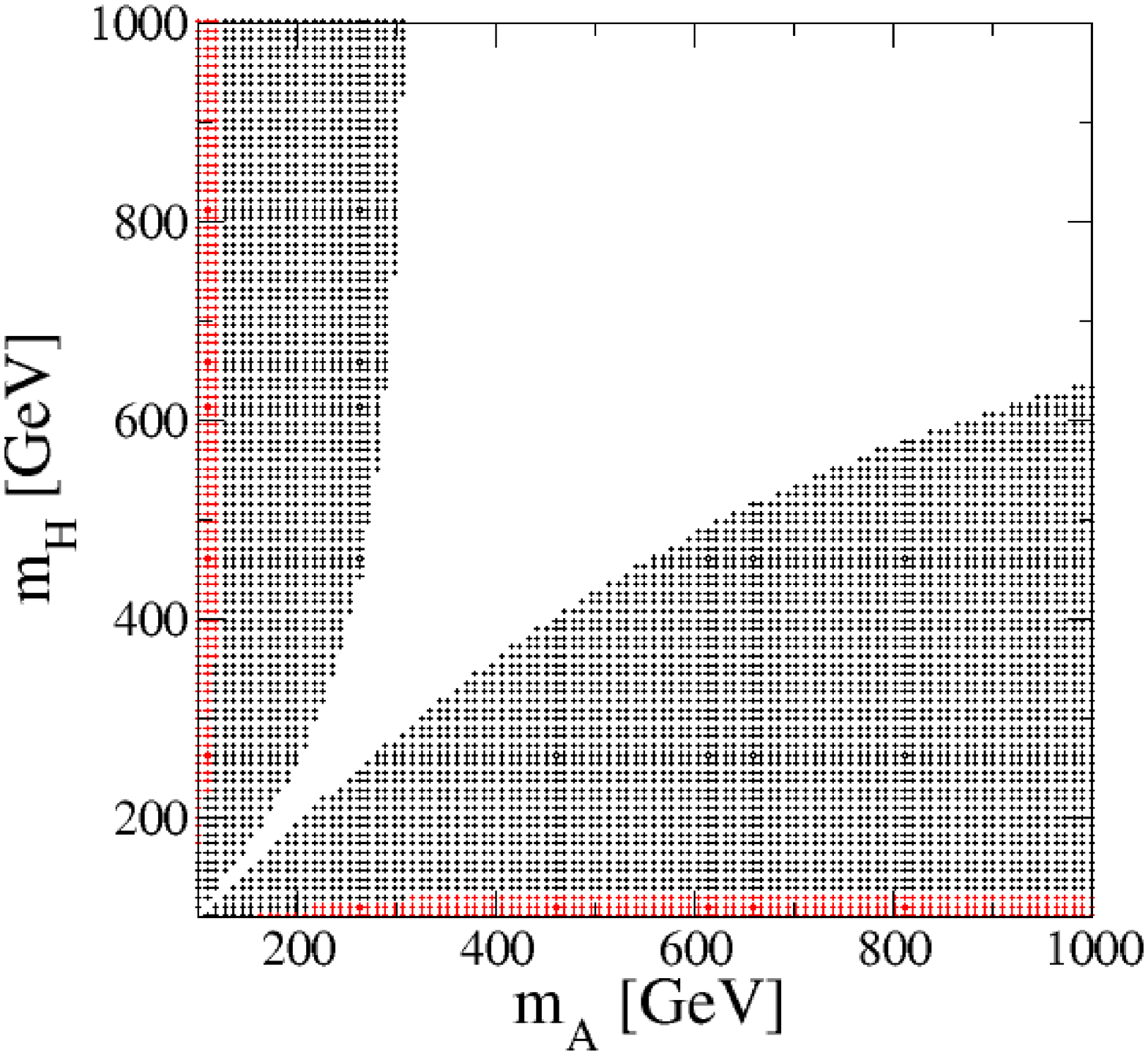} \hspace{-5mm}
\includegraphics[width=55mm]{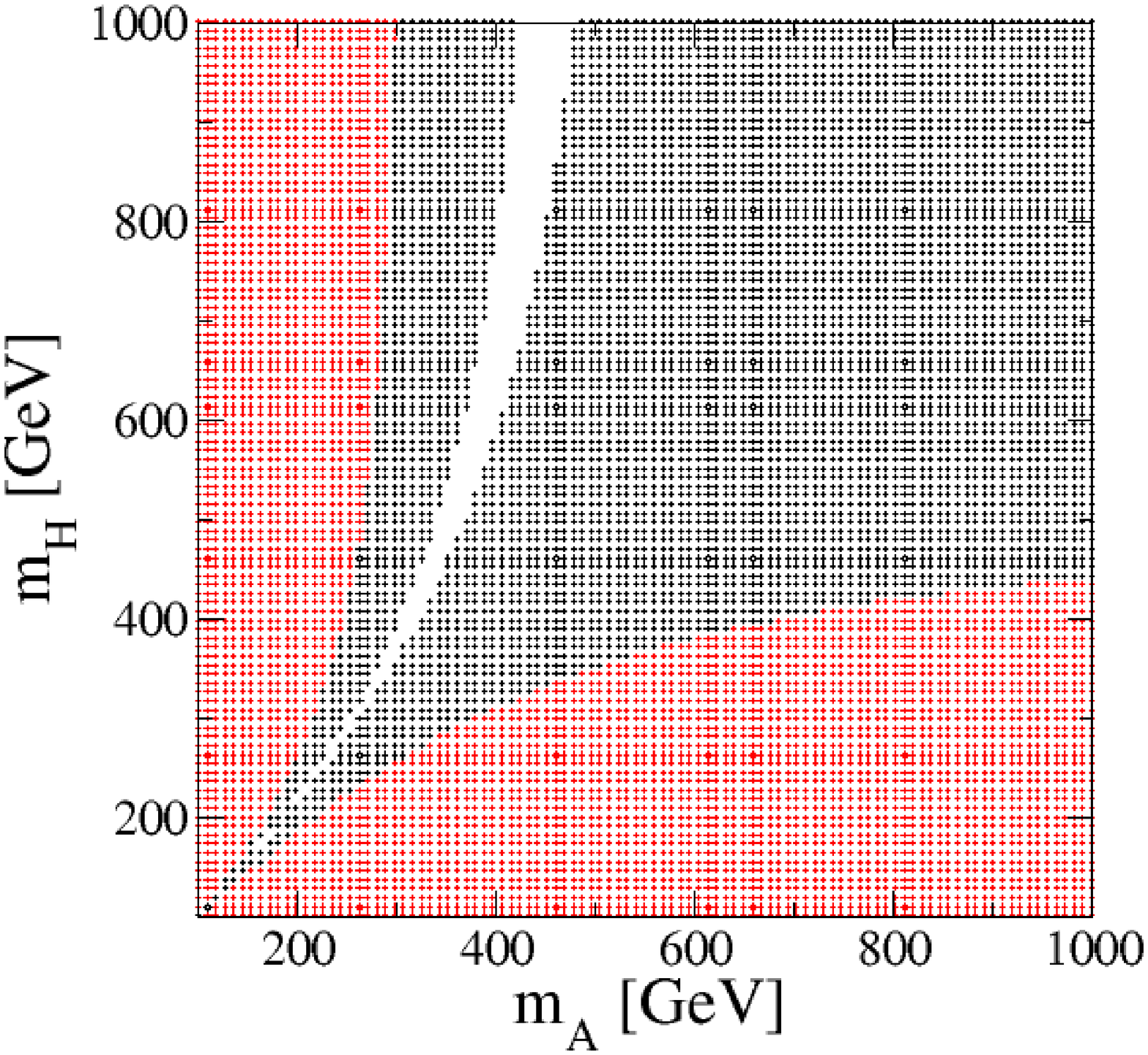} \\
\includegraphics[width=55mm]{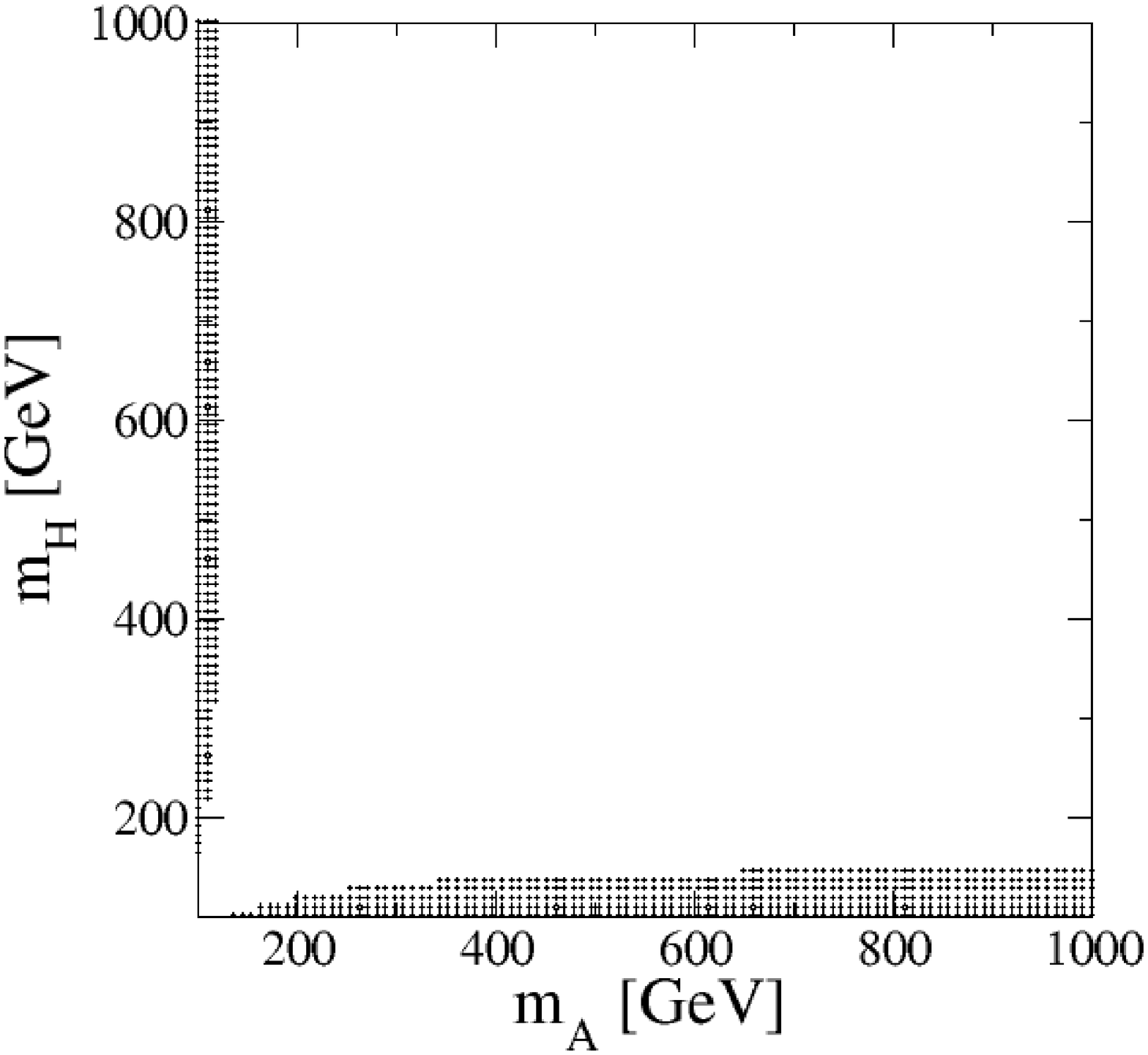}  \hspace{-5mm}
\includegraphics[width=55mm]{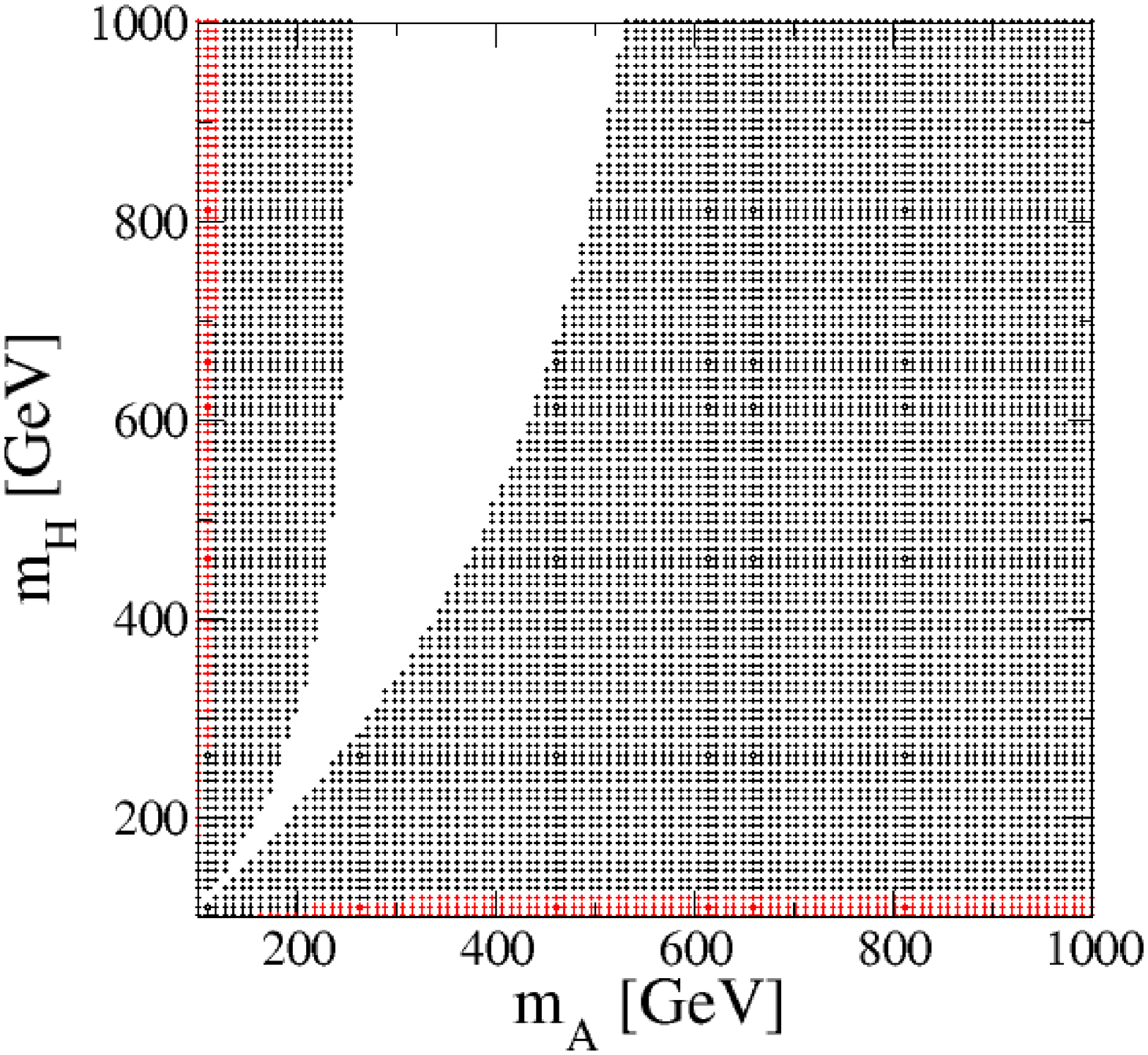} \hspace{-5mm}
\includegraphics[width=55mm]{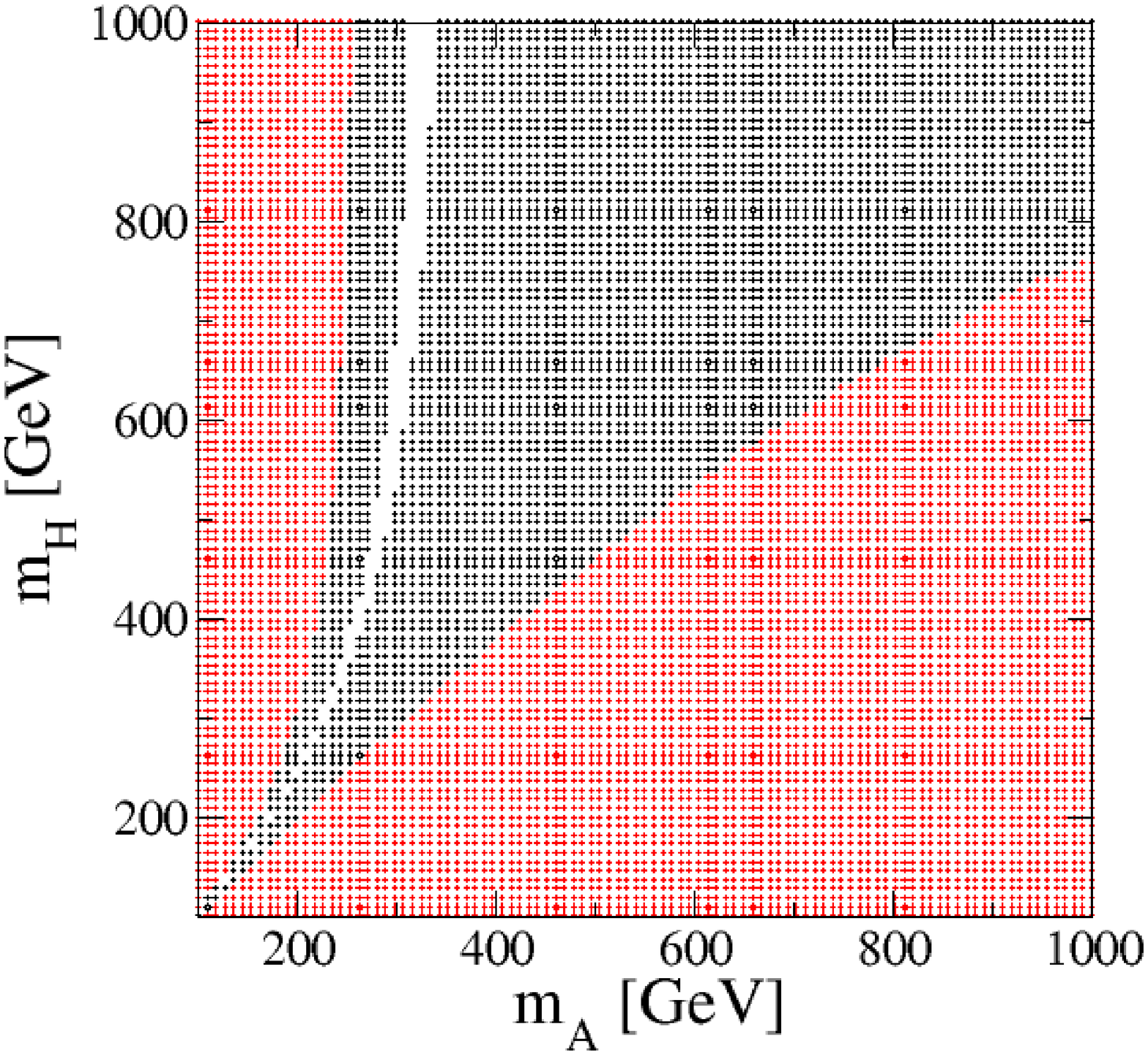}
   \caption{Constraint from the neutral meson mixings on the parameter space of $m_A$-$m_H$ in the alignment limit defined in Eq.~(\ref{align}) with 
$\sin(\beta-\alpha)=1$ (upper panels), $\sin(\beta-\alpha)=0.97$ (middle panels) and $\sin(\beta-\alpha)=0.94$ (bottom panels). 
The sign of $\cos(\beta-\alpha)$ is taken to be positive. 
The left, center and right panel show the case for $\tan\beta=3$, 10 and 30, respectively. 
The black and red shaded regions are excluded by $\Delta m_B$ and $\Delta m_D$, respectively. 
}
   \label{mixing}
\end{center}
\end{figure}

In Fig.~\ref{mixing}, we show the allowed parameter region from the meson mixing data $\Delta m_K$, $\Delta m_B$ and $\Delta m_D$. 
In these plot, we take $\tan\beta=3$ (left), $10$ (center) and 30 (right). 
The value of $\sin(\beta-\alpha)$ is taken to be 1 (upper panels), 0.97 (center panels) and 0.94 (lower panels), and the sign of $\cos(\beta-\alpha)$
is taken to be positive. We confirm that the case with  $\cos(\beta-\alpha)<0$ is almost the same as that with $\cos(\beta-\alpha)>0$. 
The black and red shaded regions are respectively excluded by 
$\Delta m_B$ and $\Delta m_D$, where in these regions, the predictions for $\Delta m_B$ and $\Delta m_D$
exceed the measured values given in Eq.~(\ref{exp}).   
We note that $\Delta m_K$ does not exclude the parameter space shown in this figure. 
As we can see that $\Delta m_B$ gives the strongest constraint, and the excluded region becomes wider when the value of $\tan\beta$ increases. 
However, in the case of $\sin(\beta-\alpha)=1$, 
the region with $m_A^{} \simeq m_H^{}$ is allowed even for the case with small masses and large $\tan\beta$, because 
the cancellation happens between the contributions from $A$ and $H$. 
Similar cancellation also happens for $\sin(\beta-\alpha)\neq 1$ among $h$, $H$ and $A$, but it does in the different regions from  $m_A^{} \simeq m_H^{}$, and 
the allowed region becomes smaller as the deviation in $\sin(\beta-\alpha)$ from unity becomes larger.

Finally, we briefly comment on flavour constraints related to the charged Higgs boson mediation such as $b\to s\gamma$~\cite{bsg,bsg2}, 
$B\to \tau\nu$~\cite{Btaunu}, and the leptonic tau decay~\cite{Tau_decay} processes. 
Because the third generation fermion couplings to $H^\pm$ have the similar structure as those in the type-II THDM, 
we expect that the similar bound on the mass of $H^\pm$ and $\tan\beta$ is obtained. 
For example, from the $b\to s\gamma$ data, we obtain the lower bound on $m_{H^\pm}$ at 95\% CL to be about 480 GeV~\cite{bsg2} in the type-II THDM with $\tan\beta\gtrsim 1$. 
The $B\to \tau\nu$ data also constrains especially a large $\tan\beta$ region. 
For example, $\tan\beta \gtrsim 30~(45)$ with $m_{H^\pm}=300$ (500) GeV is excluded at 95\% CL~\cite{Stal}. 
The comprehensive study on the constraint from the flavour experiments have been done in Refs.~\cite{Stal,Watanabe,Eber} in a $Z_2$ symmetric version of the THDMs. 

\section{Higgs Phenomenology}

In this section, we discuss the phenomenology of Higgs bosons. 
We take the limit of $v^2/u^2\to 0$, where the extra gauge bosons and exotic quarks are decoupled from the theory, and 
the scalar sector effectively becomes the THDM, i.e., we have $h$, $H$, $A$ and $H^\pm$ as the physical degrees of freedom as mentioned in Sec.~II-C.
In this case, the alignment limit of the mass matrix of the CP-even Higgs bosons is naturally realized as seen in Eq.~(\ref{align2}), 
so that we can safely take the masses of the Higgs bosons to be 
${\cal O}(100)$ GeV without conflicting with the flavour constraints as we discussed in Sec.~III-B. 
 
We first consider the phenomenology regarding to the SM-like Higgs boson $h$, and then that 
to the extra Higgs bosons $H$, $A$ and $H^\pm$. 
The relevant trilinear Higgs boson couplings are given as follows
\begin{align}
{\cal L}_{\text{int}} &= 
\frac{2m_W^2}{v}(hs_{\beta-\alpha} + H c_{\beta-\alpha}) W_\mu^+W^{-\mu}
+ \frac{m_Z^2}{v}(hs_{\beta-\alpha}+ Hc_{\beta-\alpha} ) Z_\mu Z^{\mu} \notag\\
&+ \frac{1}{v}\overline{q_L}\, \Gamma_q^h \, q_R h
 + \frac{1}{v}\overline{q_L}\, \Gamma_q^H \, q_R H
+\frac{\sqrt{2}}{v}\left[ \overline{u_L} \,\Gamma_d^{H^\pm} \,d_R  +
\overline{u_R}\,(\Gamma_u^{H^\pm})^\dagger \,d_L \right]H^+ + \text{h.c.} \notag\\
&+  \frac{m_{e}}{v}\bar{e}_Le_R (\xi_h h + \xi_H H +it_\beta A )
+\frac{\sqrt{2}m_{e}}{v}\bar{\nu}_L e_R H^+ + \text{h.c.}, \label{yint2}
\end{align}
where 
\begin{align}
\Gamma_d^{h}&=V_L^d \text{diag}(\zeta_h,\zeta_h,\xi_h) (V_L^d)^\dagger M_d^{\text{diag}}, \quad
\Gamma_u^{h}=V_L^u\, \text{diag}(\xi_h,\xi_h,\zeta_h) (V_L^u)^\dagger M_u^{\text{diag}},    \label{Gam_h}\\
\Gamma_d^{H}&=V_L^d \text{diag}(\zeta_H,\zeta_H,\xi_H) (V_L^d)^\dagger M_d^{\text{diag}}, \quad
\Gamma_u^{H}=V_L^u\, \text{diag}(\xi_H,\xi_H,\zeta_H) (V_L^u)^\dagger M_u^{\text{diag}}, 
\end{align}
with
\begin{align}
\zeta_h &= \frac{c_\alpha}{s_\beta} = s_{\beta-\alpha}+\frac{1}{t_\beta} c_{\beta-\alpha}, \quad
\xi_h = -\frac{s_\alpha}{c_\beta} = s_{\beta-\alpha}-t_\beta c_{\beta-\alpha},  \label{xih}\\
\zeta_H &= \frac{s_\alpha}{s_\beta} = \frac{1}{t_\beta}s_{\beta-\alpha}+ c_{\beta-\alpha}, \quad
\xi_H =    \frac{c_\alpha}{c_\beta} = -t_\beta s_{\beta-\alpha}+ c_{\beta-\alpha}. 
\end{align}
In Eq.~(\ref{yint2}), we omitted the flavour index for the Yukawa interaction. 
We can see that when we take $\sin(\beta-\alpha)=1$, all the coupling constants of $h$ become the same as 
the corresponding SM Higgs boson couplings. 
On the other hand, the $HVV$ $(V=W,Z)$ couplings vanish in this limit, but the Yukawa couplings for $H$ do not. 
Thus, $H$ has a fermiophilic nature in this case as it is also seen in $A$. 

\subsection{Phenomenology for the SM-like Higgs boson}

We focus on the deviation in the $h$ couplings from the SM prediction. 
In extended Higgs sectors, in general, 
the $h$ couplings deviate from the SM predictions, because of the mixing between $h$ and extra Higgs bosons, and also the mixing among VEVs of Higgs multiplets. 
The important point is that the pattern of the deviation strongly depends on the structure of the Higgs sector.  
Therefore, we can determine the structure of the Higgs sector by identifying the pattern of deviation 
in the $h$ couplings measured at collider experiments. 
Precise measurements of the $h$ couplings will be done at future collider experiments such as High-Luminosity LHC~\cite{HLLHC_ATLAS,HLLHC_CMS} 
and the International Linear Collider (ILC)~\cite{ILC-white}. 
In Refs.~\cite{FP}, the deviations in the Higgs boson couplings have been discussed at the tree level 
in various extended Higgs sectors such as THDMs and models with extra isospin singlets, 
triplets and septets which satisfy the electroweak $\rho$ parameter being unity at the tree level. 
It has been clarified that these models can be discriminated by using the deviations in $hVV$ and $hff$ couplings. 
Radiative corrections to the $h$ couplings have also been studied in THDMs~\cite{FP-THDM}, a model with a singlet~\cite{FP-HSM} and that with a triplet~\cite{FP-HTM}. 

\begin{table}[t]
\begin{center}
{\renewcommand\arraystretch{1}
\begin{tabular}{c|cccc}\hline\hline
Type & $\kappa_{u^i}$ & $\kappa_{d^i}$ & $\kappa_{e^i}$ & $\kappa_V^{}$ \\\hline
type-I &$\zeta_h$ &$\zeta_h$ &$\zeta_h$&$\sin(\beta-\alpha)$ \\\hline
type-II &$\zeta_h$&$\xi_h$&$\xi_h$&$\sin(\beta-\alpha)$ \\\hline
type-X &$\zeta_h$ &$\zeta_h$&$\xi_h$&$\sin(\beta-\alpha)$ \\\hline
type-Y &$\zeta_h$ &$\xi_h$&$\zeta_h$&$\sin(\beta-\alpha)$ \\\hline\hline
\end{tabular}}
\caption{The scaling factors in the THDMs with a softly-broken $Z_2$ symmetry. }
\label{xi}
\end{center}
\end{table}

In our model, the $h$ couplings deviate from the SM prediction in the case of $\sin(\beta-\alpha)\neq 1$ at the tree level 
which corresponds to the case with a non-zero deviation in the $hVV$ couplings
as it is seen in Eq.~(\ref{yint2}). 
The pattern of the deviation in the Yukawa couplings for the third generation lepton (quarks) is exactly (almost) the same as that in the type-II THDM at the tree level. 
However, the difference in the prediction from the type-II THDM appears 
in the correlation between the deviation in the $h$ coupling with the second and the third generation quarks. 
In fact, it is seen in Eq.~(\ref{Gam_h}) that 
the (3,3) and (2,2) element of the coupling matrix $\Gamma_q^h$ are almost\footnote{The meaning of almost here is that, for instance, 
the (3,3) element of $\Gamma_d^h$ is not exactly determined by $\xi_h$, i.e.,
the $\zeta_h$ dependence also enters,  due to the small off-diagonal elements of $V_L^d$. }
determined by the different valuable $\xi_h$ or $\zeta_h$ defined in Eq.~(\ref{xih}). 

\begin{figure}[t]
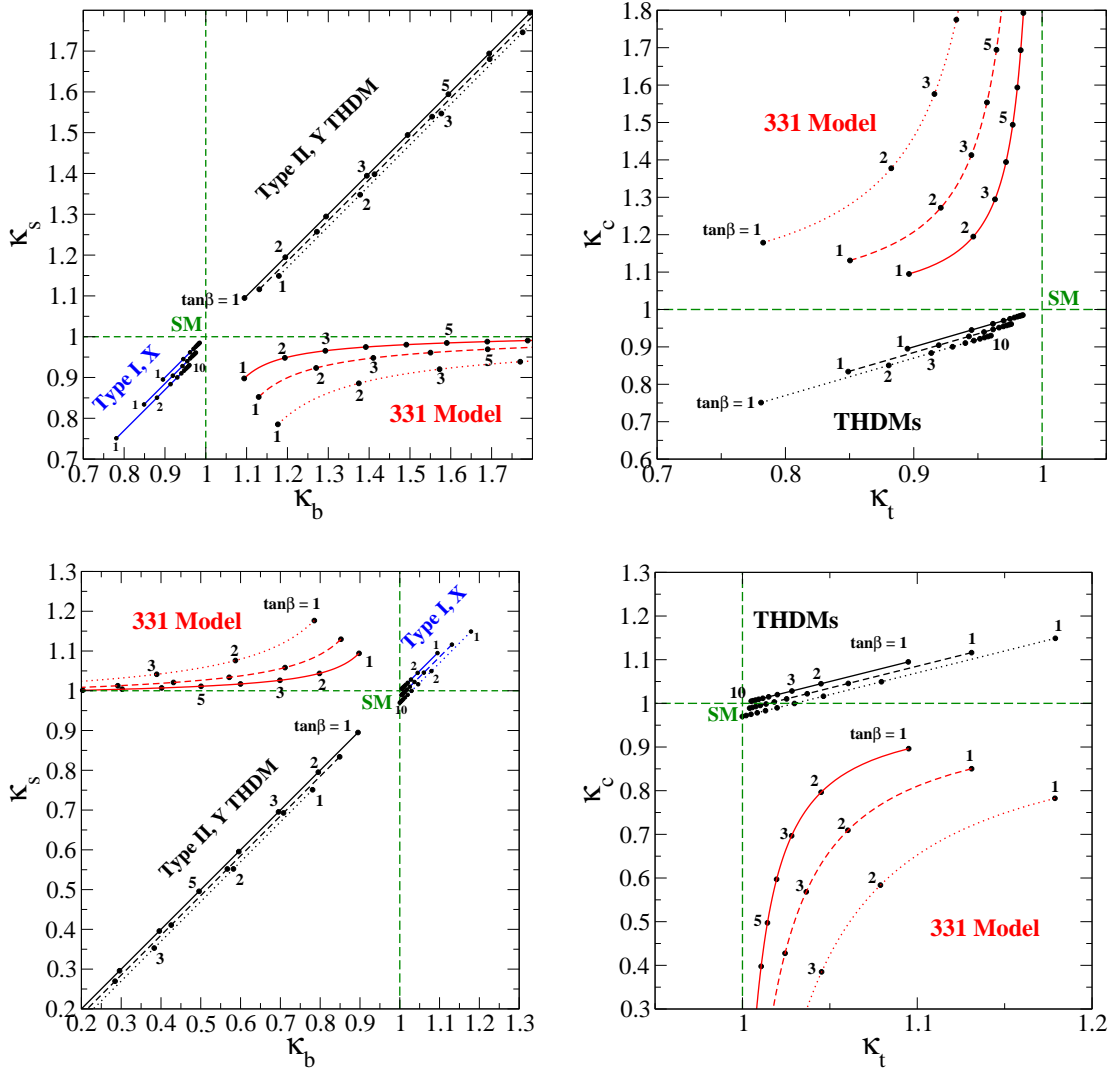

\begin{center}
 \includegraphics[width=70mm]{Kb-Ks_minus.eps}\hspace{5mm}
 \includegraphics[width=70mm]{Kt-Kc_minus.eps} \\\vspace{5mm}
 \includegraphics[width=70mm]{Kb-Ks.eps}\hspace{5mm}
 \includegraphics[width=70mm]{Kt-Kc.eps}
   \caption{Predictions for the scaling factor of the Yukawa couplings in our model and the THDMs with a softly-broken $Z_2$ symmetry. 
The upper (lower)-left and upper (lower)-right panel show the prediction on the $\kappa_b$-$\kappa_s$ and $\kappa_t$-$\kappa_c$ plane
with $\cos(\beta-\alpha)<0$ $(\cos(\beta-\alpha)>0)$, respectively.
The solid, dashed and dotted curve in each panel show the case with $\sin(\beta-\alpha)=0.995$, 0.99 and 0.98, respectively. 
The value of $\tan\beta$ is varied from 1 to 10, and the each dot on the curves shows the prediction with a specific value of $\tan\beta$.  }
\label{kappa}
\end{center}
\end{figure}

In order to see the correlation between the second and the third quark Yukawa coupling of $h$, we define the scaling factor as 
\begin{align}
\kappa_{f^i} \equiv \frac{\text{Re}[c_{hf^if^i}^{\text{331 Model}}] }{\text{Re}[c_{hf^if^i}^{\text{SM}}]}, \quad
\kappa_V^{}  \equiv \frac{\text{Re}[c_{hVV}^{\text{331 Model}}] }{\text{Re}[c_{hVV}^{\text{SM}}]}, \quad V=W,~Z,  
\end{align}
where $c_{hf^if^i}^{\text{SM}}$ and $c_{hVV}^{\text{SM}}$ ($c_{hf^if^i}^{\text{331 Model}}$ and $c_{hVV}^{\text{331 Model}}$) are 
the $h\bar{f^i}f^i$ and $hVV$ coupling in the SM (our model), respectively. 
To clearly show the flavour dependence, we keep the flavour index $i$ in the above expressions. 
From Eq.~(\ref{yint2}), these scaling factors are calculated as 
\begin{align}
\kappa_{d^i}  & = \frac{(\Gamma_d^h)_{ii},}{m_{d^i}},~~\kappa_{u^i}  = \frac{(\Gamma_u^h)_{ii},}{m_{u^i}},~~
\kappa_{e^i}    = \xi_h, ~~
\kappa_V^{} = \sin(\beta-\alpha). 
\end{align}
It is important to comment on the scaling factors in the 
THDMs with a softly-broken $Z_2$ symmetry, where the type-II THDM is the one which has the same structure of the Yukawa interaction as that of the minimal supersymmetric SM. 
In addition to the type-II model, we can define the other three independent types of the THDMs, the so-called type-I, type-X and type-Y~\cite{AKKY}. 
The scaling factors for the Yukawa couplings are flavour universal in the THDMs, and these formulae are given in Table~\ref{xi}. 

In Fig.~\ref{kappa}, we show the correlation of the scaling factors $\kappa_b$ and $\kappa_s$ (left panel),
and $\kappa_t$ and $\kappa_c$ (right panel) in our model and in the THDMs. 
The upper and lower panels respectively show the case of $\cos(\beta-\alpha)<0$ and $\cos(\beta-\alpha)>0$. 
In each panel, the dots on the curves show the prediction in the different value of $\tan\beta$, where the
interval of each dot corresponds to the one value difference in $\tan\beta$. 
The solid, dashed and dotted curves show the case for $\sin(\beta-\alpha)=0.995$, 0.99 and 0.98, respectively, where 
these correspond to the case with $0.5$, 1 and 2\% deviation in the $hVV$ couplings. 
For the predictions in the THDMs, we slightly shift the three curves from their original locations in order to clearly show the three cases. 
We can clearly see that the predictions in our model and in the THDMs are given in the different region on the both 
$\kappa_b$-$\kappa_s$ and the $\kappa_t$-$\kappa_c$ plane. 
Therefore, we can distinguish our model from the THDMs from the precise measurement of the Yukawa couplings as long as $\kappa_V^{}\neq 1$ is given. 
We note that the four types of the THDMs are also distinguished by looking at the correlation among $\kappa_b$, $\kappa_\tau$ and $\kappa_t$ as shown in Ref.~\cite{FP}.

\subsection{Phenomenology for the extra Higgs bosons}

We discuss the phenomenology of the extra Higgs bosons in this subsection, i.e., 
we first calculate the decay branching fractions and then evaluate the production cross sections at the LHC. 

Basically, the decay property of $H$, $A$ and $H^\pm$ is similar to the corresponding extra Higgs boson in the THDMs in the context that 
they mainly decay into a fermion pair when we take $\sin(\beta-\alpha)=1$. 
If there is a non-zero mass difference among the extra Higgs bosons, the decay associated with a weak boson can also be dominant such as 
$H^\pm \to A W^\pm/H W^\pm$ if $m_{H^\pm}> m_{A/H}^{}$. 
The most important decay property in our model is seen in  
the flavour violating decay modes of the extra Higgs bosons
which are naturally suppressed in the THDMs.  
When $\sin(\beta-\alpha)\neq 1$ is given, the fermiophilic nature of $H$ is lost, and then 
the decay modes with the $W^+W^-$ and $ZZ$ become important. 
Besides, the $H\to hh$ decay mode also opens, because the $Hhh$ coupling is proportional to $\cos(\beta-\alpha)$ as given in Eq.~(\ref{Hhh}). 
These features with $\sin(\beta-\alpha)\neq 1$ are also seen in the THDMs. 
From the above discussion, the characteristic decay mode, i.e., the flavour violating processes, is clearly seen in $\sin(\beta-\alpha)\simeq 1$. 

In the following, we numerically show the decay branching fractions of $H$, $A$ and $H^\pm$ in the case of $\sin(\beta-\alpha)=1$. 
In this analysis, we use the following SM input parameter~\cite{PDG}:
\begin{align}
&m_t = 173.21~\text{GeV},~~\bar{m}_b = 3.0~\text{GeV},~~\bar{m}_c = 0.677~\text{GeV},~~\bar{m}_s = 0.0934~\text{GeV},\notag\\
&m_Z=91.1876~\text{GeV},~m_W=80.385~\text{GeV},~G_F=1.1663787\times 10^{-5}~\text{GeV}^{-2},\notag\\
&m_h = 125~\text{GeV},~m_\tau = 1.77684~\text{GeV},~\alpha_s = 0.1185. 
\end{align}
The running quark masses $\bar{m}_b$, $\bar{m}_c$ and $\bar{m}_s$ are taken at the $m_Z$ scale~\cite{Koide}.  
We use the same values of the quark mixing matrix elements as given in Eq.~(\ref{qmatrix}). 
We note that for the neutral Higgs decays, the decay rates of $A/H \to \bar{q}_i q_j$ and $A/H \to \bar{q}_j q_i$ ($i\neq j$) are summed. 
All the relevant formulae of the decay rates of the Higgs bosons are presented in App.~C.  

\begin{figure}[t]
\begin{center}
 \includegraphics[width=70mm]{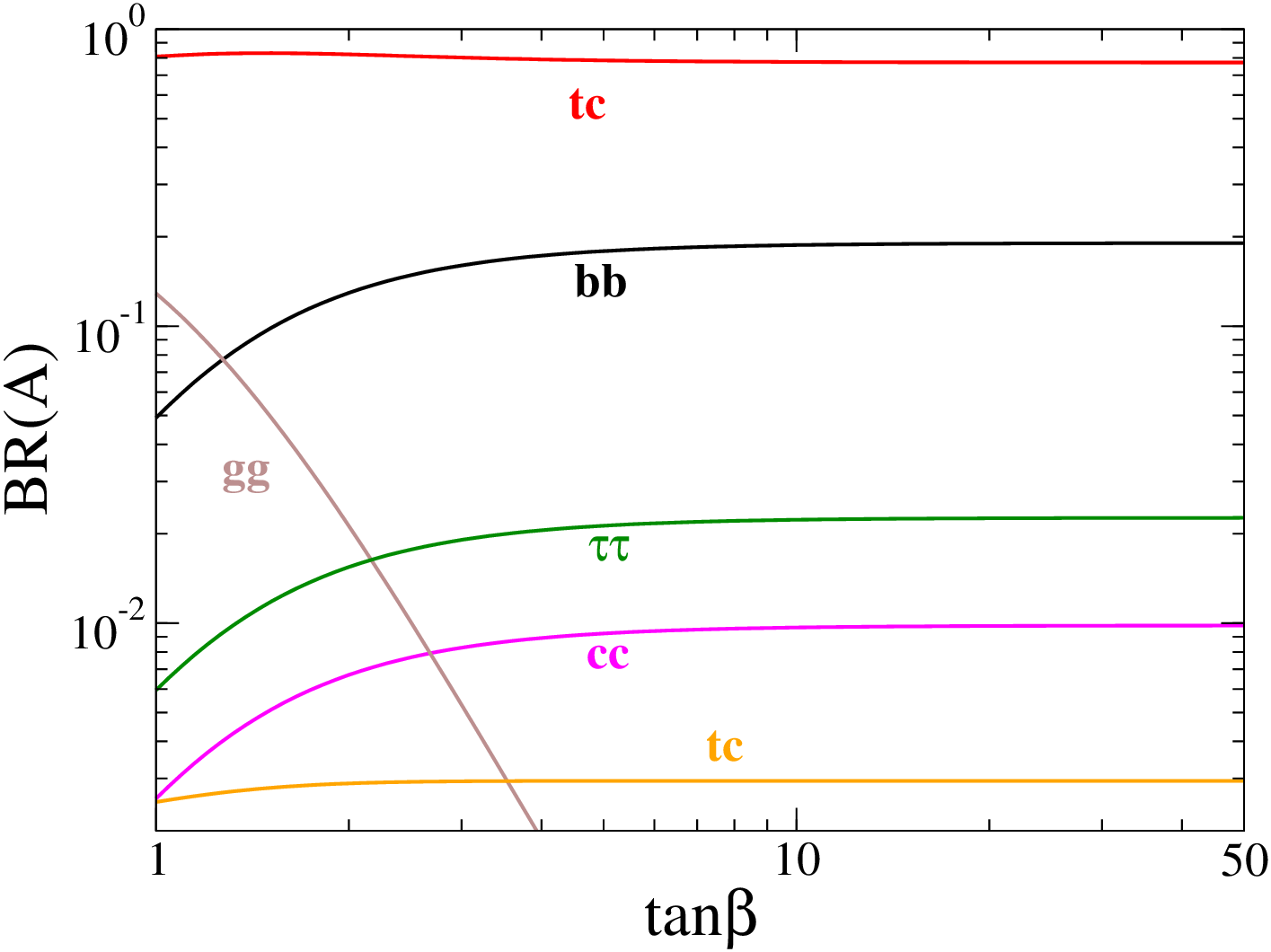}\hspace{5mm}
 \includegraphics[width=70mm]{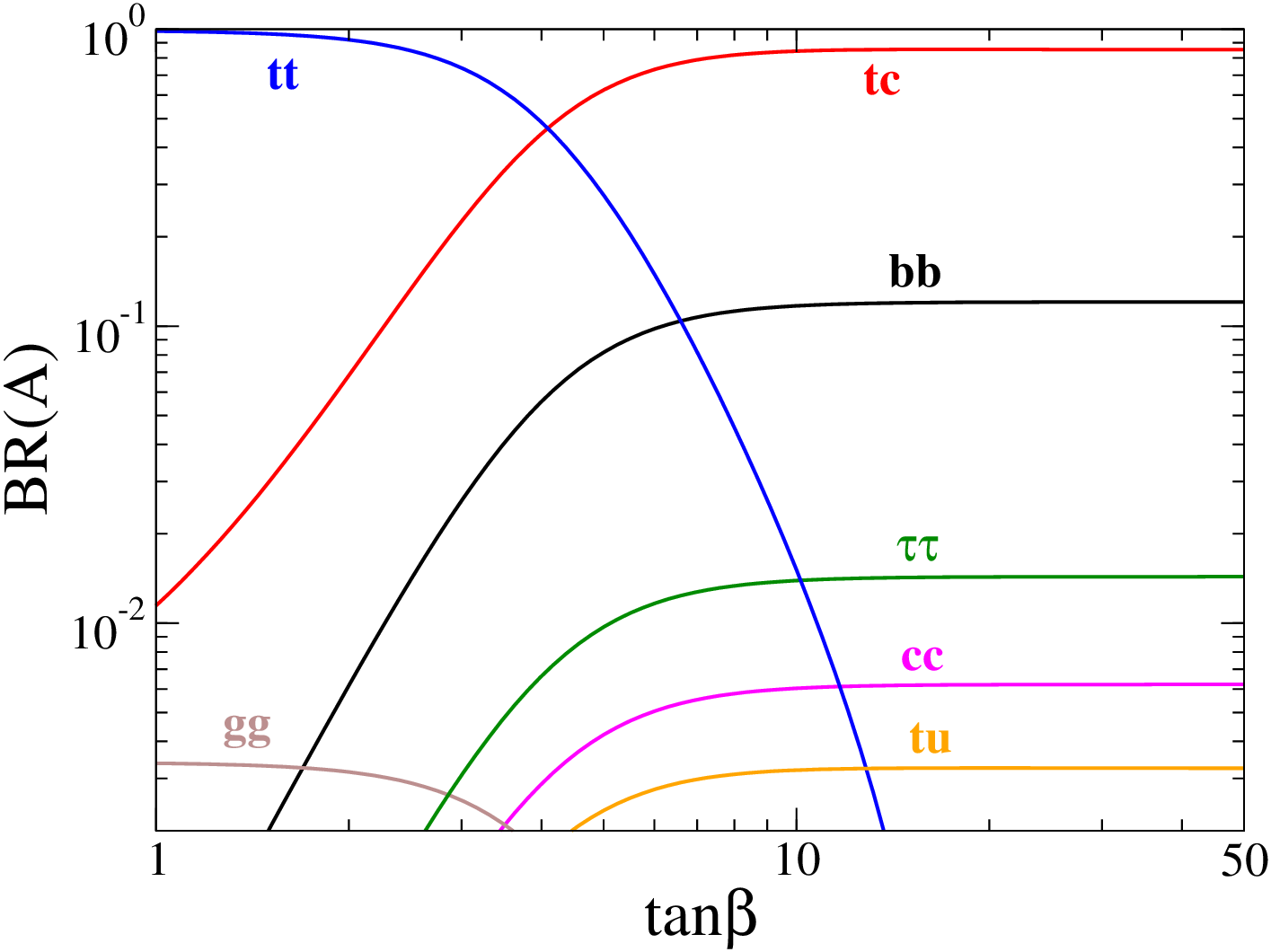}
   \caption{Decay branching fractions of $A$ as a function of $\tan\beta$. 
We take $m_A=m_H=m_{H^\pm}=300$ (500) GeV for the left (right) panel. 
}
   \label{BR_A}
\end{center}\vspace{5mm}
\begin{center}
 \includegraphics[width=70mm]{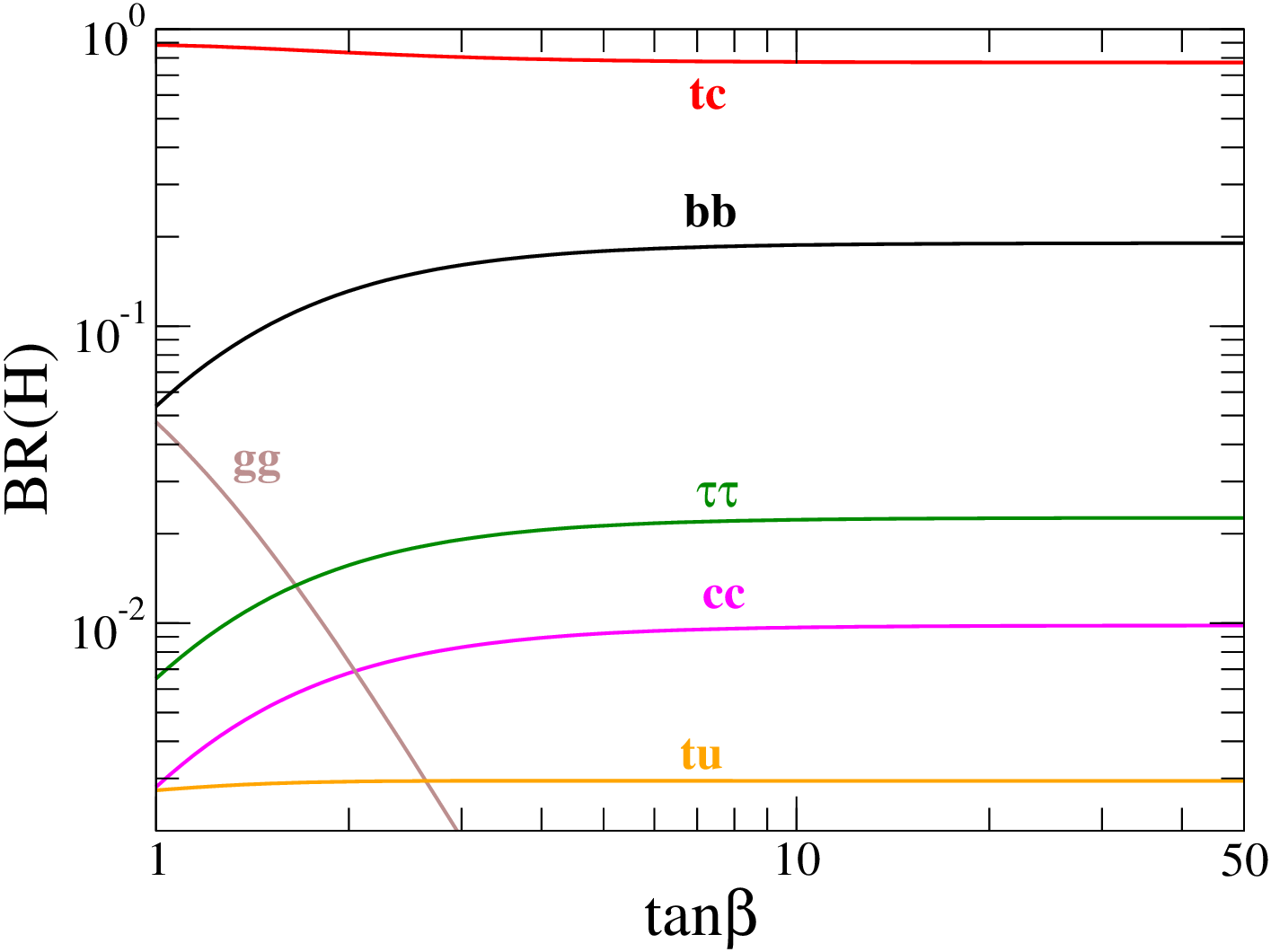}\hspace{5mm}
 \includegraphics[width=70mm]{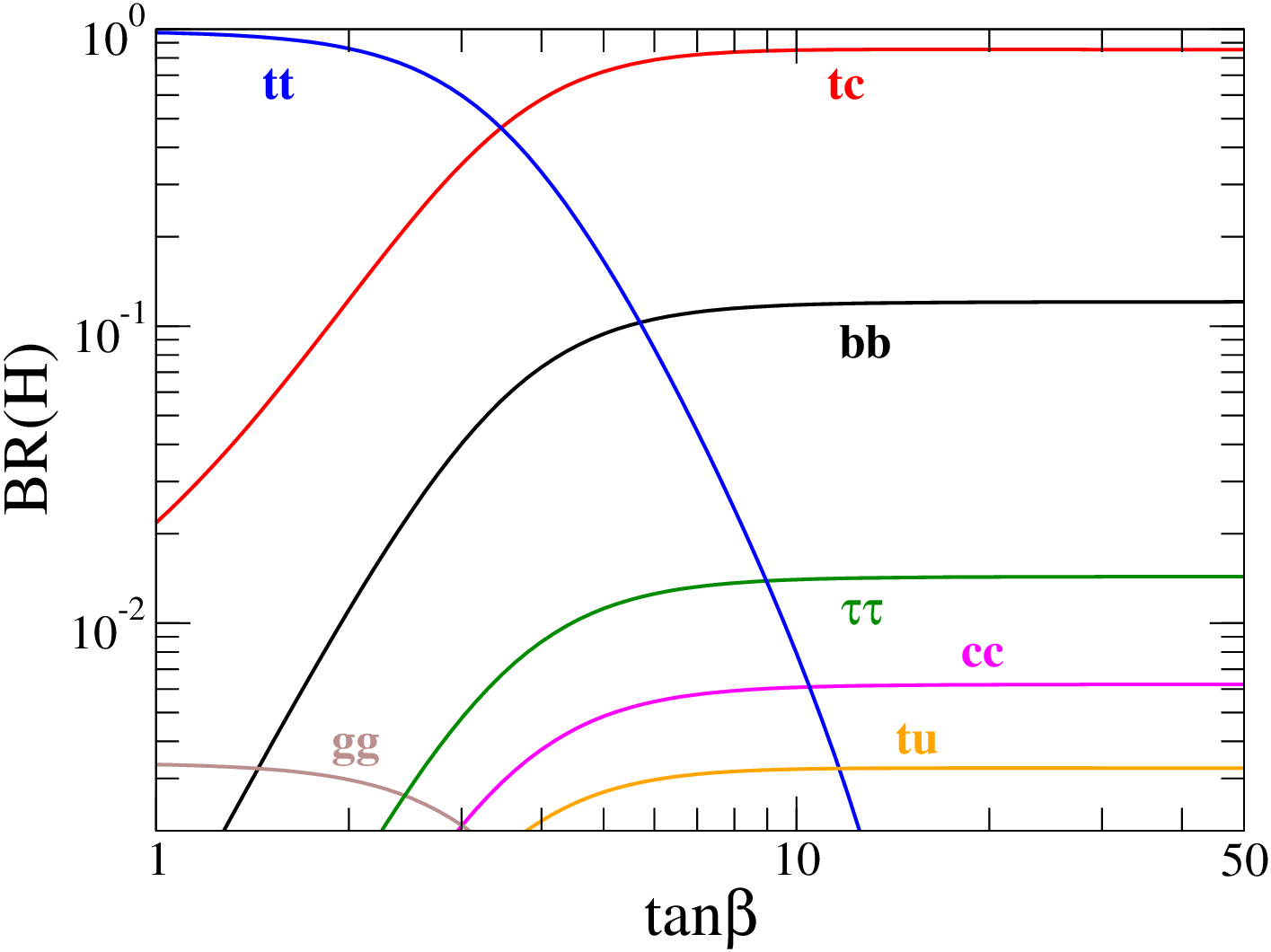}
   \caption{Decay branching fractions of $H$ as a function of $\tan\beta$. 
We take $m_A=m_H=m_{H^\pm}=300$ (500) GeV for the left (right) panel. }
   \label{BR_H}
\end{center}
\end{figure}

In Figs.~\ref{BR_A} and \ref{BR_H}, we show the decay branching fractions of $A$ and $H$ as a function of $\tan\beta$, respectively. 
The left (right) panel shows the case for $m_A=m_H=m_{H^\pm}=300$ (500) GeV. 
For the left case, we see that the $tc$ and $bb$ modes are dominant in the wide range of $\tan\beta$, where 
the former and latter mode have the branching fraction of about 80\% and about 20\%, respectively. 
Except for the small difference in the $A\to gg$ and $H\to gg$ modes, 
the branching fractions of $A$ and $H$ are almost the same. 
For the 500 GeV case shown in the right panel, 
the $t\bar{t}$ channel is kinematically allowed and this can be dominant in the small $\tan\beta$ region. 
However, when $\tan\beta \gtrsim 4$, the main decay mode is replaced by the $tc$ mode. 
We here comment on the one-loop induced decay modes of $A/H \to \gamma\gamma$ and $A/H \to Z\gamma$. 
Typically, the branching fractions of these modes are the order of $10^{-4}$-$10^{-5}$ when $m_H^{}=m_A^{}=300$ GeV.
Smaller values of the branching fractions are obtained when $\tan\beta$ and/or the masses of $A$ and $H$ increase. 

\begin{figure}[t]
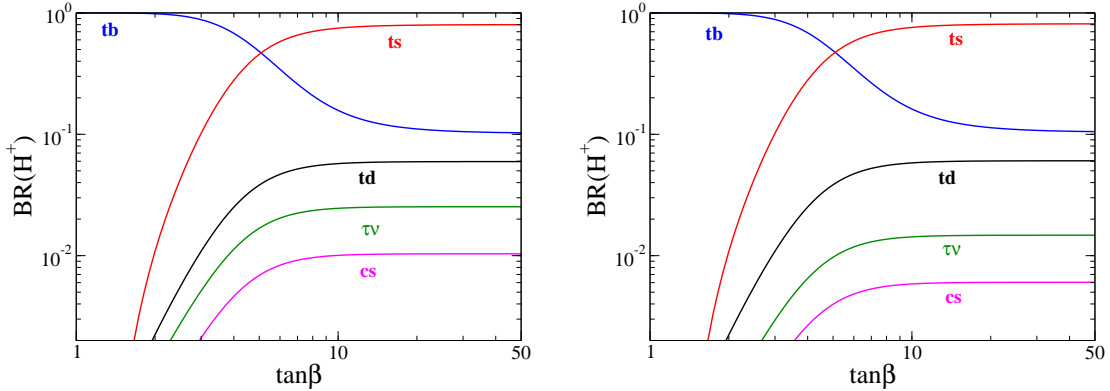

\begin{center}
 \includegraphics[width=70mm]{BR_Hp_300.eps}\hspace{5mm}
 \includegraphics[width=70mm]{BR_Hp_500.eps}
   \caption{Decay branching fractions of $H^\pm$ as a function of $\tan\beta$. 
We take $m_A=m_H=m_{H^\pm}=300$ (500) GeV for the left (right) panel. }
   \label{BR_Hp}
\end{center}
\end{figure}

In Fig.~\ref{BR_Hp}, we show the decay branching fractions of $H^+$ as a function of $\tan\beta$. 
Similar to the case for the neutral Higgs bosons, we take $m_A=m_H=m_{H^\pm}=300$ (500) GeV for the left (right) panel. 
We see that the main decay mode is changed from the $t\bar{b}$ mode to the $t\bar{s}$ mode at $\tan\beta\simeq 5$ for the 
both 300 GeV and 500 GeV case. 
These flavour violating decays $A/H\to tc$ and $H^\pm \to ts$ cannot be dominant in the four types of THDMs, so that 
these decay processes can be useful to identify our model. 

Finally, we calculate the production cross sections of the extra Higgs bosons at the LHC. 
The neutral Higgs bosons $A$ and $H$ are mainly produced via the gluon fusion mechanism: $gg \to A/H$. 
The production cross section is calculated by 
\begin{align}
\sigma(gg\to A/H) = \sigma (gg \to h_\text{SM})\times \frac{\Gamma(A/H\to gg)}{\Gamma(h_{\text{SM}}\to gg)}, 
\end{align} 
where $h_{\text{SM}}$ is the SM Higgs boson.
The analytic expression for the decay rate $\Gamma(A/H\to gg)$ into the two gluons is given in Eq.~(\ref{gg}). 
$\sigma (gg \to h_\text{SM})$ is the gluon fusion cross section for $h_{\text{SM}}$, where the mass of $h_{\text{SM}}$
is taken here to be the same as that of $A$ or $H$. 
We quote the value of $\sigma (gg \to h_\text{SM})$ at the next-to-next-to leading order in QCD from~\cite{SM-cross}. 
In addition to the gluon fusion process, the bottom quark associated production of $A$ and $H$: $gg \to b\bar{b}A/b\bar{b}H$ can also be important. 
This cross section is proportional to $|(\Gamma_b^{A/H})_{33}|^2$ which is roughly determined by $ (m_b\times\tan\beta)^2$ when $\sin(\beta-\alpha)=1$. 
Therefore, for the large $\tan\beta$ region, this production mechanism becomes important. 

\begin{figure}[t]
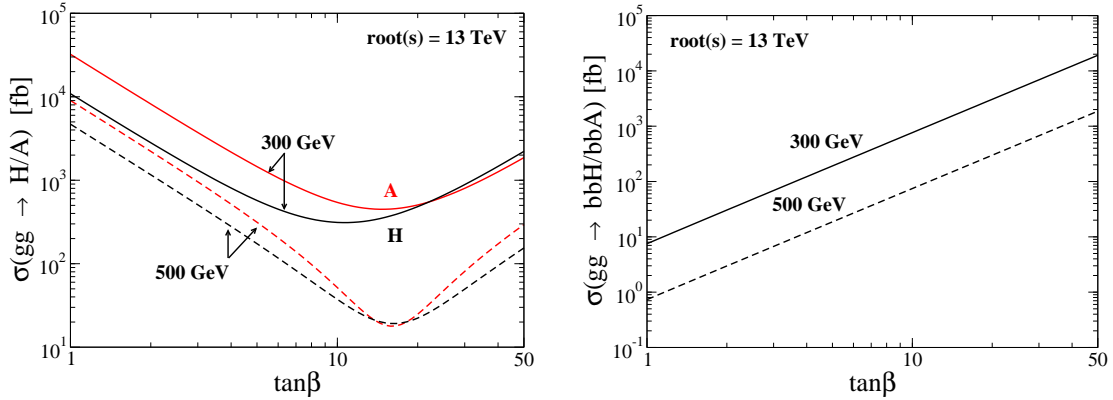

\begin{center}
 \includegraphics[width=70mm]{ggF.eps}\hspace{5mm}
 \includegraphics[width=70mm]{bas.eps}
   \caption{(Left) Production cross section of the gluon fusion process for $H$ (black) and $A$ (red) as a function of $\tan\beta$. 
(Right) Production cross section of the bottom quark associated process for $H$ or $A$  as a function of $\tan\beta$. 
For the both panels, the solid (dashed) curves show the case for the mass of $A$ or $H$ to be 300 (500) GeV, and 
the collision energy is taken to be 13 TeV. }
   \label{ggF}
\end{center}
\end{figure}

In Fig.~\ref{ggF}, we plot the production cross section for $A$ and $H$ as a function of $\tan\beta$ from the gluon fusion (left) and 
the bottom quark associated process (right) at the center of mass energy of 13 TeV. 
We use {\tt CalcHEP}~\cite{calchep} for the calculation of the bottom quark associated process, and apply to {\tt CTEQ6L}~\cite{PDF} for the parton distribution functions (PDFs).  
We separately show the gluon fusion cross section for $A$ and $H$, but we do not for the bottom quark associated process, since the cross section of 
$gg \to b\bar{b}A$ and $gg\to b\bar{b}H$ are almost the same in this configuration. 
For each process, we show the case with the masses of $A$ and $H$ to be 300 GeV (solid curve) and 500 GeV (dashed curve). 
We see that for the low $\tan\beta$ region, the gluon fusion process gives the much larger cross section as compared to the bottom quark associated process, e.g.,
at $\tan\beta\simeq 1$, the cross section is about 30 pb (10 pb) and 1 pb (0.5 pb) for $A$ ($H$) at $m_A^{}(m_H^{})=300$ and 500 GeV, respectively. 
However, this becomes small as $\tan\beta$ increases, and at around $\tan\beta = 10$, it takes the minimal value to be about 1 pb (10 fb) for the case with 
$m_A$ and $m_H$ being 300 (500) GeV. 
This is simply because the reduction of the top Yukawa coupling $(\Gamma_t^{A/H})_{33}$ whose magnitude is roughly determined by $m_t \times \cot\beta$. 
For $\tan\beta \gtrsim 10$, the bottom quark associated process gives the larger cross section as compared to the gluon fusion process. 

\begin{figure}[t]
\begin{center}
 \includegraphics[width=100mm]{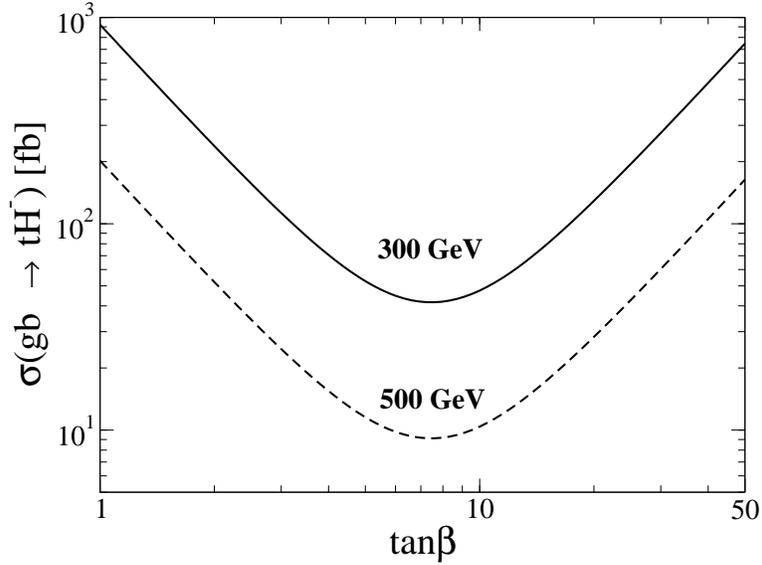}
   \caption{Cross section for the $gb\to tH^-$ process as a function of $\tan\beta$ at the collision energy of 13 TeV. 
The solid (dahsed) curves show the case for $m_{H^\pm}=300$ (500) GeV. }
   \label{Hp}
\end{center}
\end{figure}

Finally, we discuss the production of $H^\pm$ at the LHC. 
The main production mode has been known to be the gluon-bottom fusion process, i.e., $gb\to H^- t$~\cite{HHG,gb} when the mass of charged Higgs bosons is larger than the 
top quark mass. 
We calculate the production cross section by using {\tt CalcHEP} with {\tt CTEQ6L} for PDFs as it was done in the calculation of the cross section of the 
bottom quark associated production. 
In Fig.~\ref{Hp}, we show the cross section of the $gb\to H^- t$ process as a function of $\tan\beta$ in the case of $m_{H^\pm}=300$ GeV (solid curve) and 500 GeV (dashed curve). 
Similar to the gluon fusion process, the cross section becomes minimum at around $\tan\beta=10$, while it gives large values
at the low and high $\tan\beta$ case, e.g., 
we obtain 0.9 (0.2) pb at $\tan\beta\simeq 1$ for $m_{H^\pm}=300$ (500) GeV, and
0.7 (0.15) pb at $\tan\beta\simeq 50$ for $m_{H^\pm}=300$ and (500) GeV. 

In fact, these results of the cross section of $A$, $H$ and $H^\pm$ are almost the same as those of the corresponding Higgs bosons in the type-II THDM. 
However, we expect that our model is distinguishable by using the signature of the flavour violating decays of the Higgs bosons.

\section{Conclusions}

We have discussed the phenomenology of the model based on the $SU(3)_c\times SU(3)_L\times U(1)_X$ gauge theory with the minimal form of the Higgs sector which is 
composed of the three $SU(3)_L$ triplet scalar fields. 
We have shown that our Higgs sector effectively becomes THDMs after the spontaneous symmetry breaking $SU(3)_L\times U(1)_X\to SU(2)_L\times U(1)_Y$. 
One of the most important features in our effective THDM originating from the 331 model is seen in the structure of the quark Yukawa interactions, in which 
the first and the second generation quarks couple to the different Higgs doublet from that couples to the third generation quarks. 
This flavour dependent structure inevitably causes FCNC's at the tree level via the Higgs boson mediations. 
In order to avoid the constraint from the flavour experiments, we have taken the alignment limit on the mass matrix of the CP-even Higgs bosons, which 
is naturally realized in the limit of $v^2/u^2 \to 0$. 
Under the alignment limit, we have shown that the Higgs boson masses of ${\cal O}(100)$ GeV are consistent with the considered $K^0$-$\bar{K}^0$, 
$B^0$-$\bar{B}^0$ and $D^0$-$\bar{D}^0$ mixings. 
In this allowed parameter space, we have considered
the deviation in the SM-like Higgs boson couplings from the SM  predictions. 
We have found that in the case of $\sin(\beta-\alpha)\neq 1$, our predictions on the $\kappa_b$-$\kappa_s$ and  $\kappa_t$-$\kappa_c$ plane appear 
in the region different from that in the THDMs with a softly-broken $Z_2$ symmetry. 
We can thus distinguish our model from the THDMs by looking at the deviations in these quark Yukawa couplings. 
We have also investigated the properties of the extra Higgs bosons, i.e., the decays and productions at the LHC.  
We have found that the flavour violating Higgs boson decay modes, e.g., 
$H/A \to tc$ and $H^\pm \to ts$ are dominant in the wide region of the parameter space. 
These flavour violating decays of the extra Higgs bosons can be useful to identify our model, and to discriminate our model from the THDMs in addition to 
using the deviation of the SM-like Higgs boson couplings.

\vspace*{4mm}
\section*{Acknowledgments}
H.O. expresses his sincere gratitude toward Prof. Seungwon Baek and the KIAS hospitality during his visit, as well as
all the KIAS members, Korean cordial persons, foods, culture and weather.
This work is supported by JSPS postdoctoral fellowships for research abroad (KY), 
NRF Research No. 2009-0083526 of the Republic of Korea (YO)
and the United States Department of Energy grant (DE-SC0013680) (NO).

\begin{appendix}

\section{Higgs boson couplings to weak bosons}

We give the expressions for the Higgs boson couplings with weak gauge bosons. 
The Higgs-Gauge-Gauge type interaction terms are extracted by 
\begin{align}
{\cal L}_{\text{int}} &= 
\frac{g^2v}{2}\sum_{\alpha=1-3}(c_\beta R_{1\alpha} + s_\beta R_{2\alpha})H_\alpha W_\mu^+W^{-\mu}
+\frac{g^2}{2}\sum_{\alpha=1-3}(c_\beta v R_{1\alpha} + u R_{3\alpha})H_\alpha W_\mu^{\prime +}W^{\prime- \mu}
\notag\\
&+ \frac{g^2v}{3c_{331}^2}\sum_{\alpha=1-3}\Big[c_\beta c_Z^2 R_{1\alpha} 
+\frac{s_\beta}{4}\left(c_Z^2 + 3c_{331}^2s_Z^2+2\sqrt{3}c_Z^{} s_Z^{}\right)R_{2\alpha}\notag\\
&\hspace{4.2cm}+\frac{u}{4v}\left(c_Z^2 + 3c_{331}^2s_Z^2-2\sqrt{3}c_Z^{} s_Z^{}\right)R_{3\alpha}
 \Big]H_\alpha Z_\mu Z^{\mu}\notag\\
&+ \frac{g^2v}{3c_{331}^2}\sum_{\alpha=1-3}\Big[\{c_\beta s_Z^2 R_{1\alpha} 
+\frac{s_\beta}{4}\left(3c_Z^2c_{331}^2 + s_Z^2-2\sqrt{3}c_Z^{} s_Z^{}\right)R_{2\alpha}\notag\\
&\hspace{4.2cm}+\frac{u}{4v}\left(3c_Z^2c_{331}^2 + s_Z^2+2\sqrt{3}c_Z^{} s_Z^{}\right)R_{3\alpha}
 \Big]H_\alpha Z_\mu' Z^{\prime\mu}\notag\\
&+ \frac{g^2v}{3c_{331}^2}\sum_{\alpha=1-3}\Big\{-2c_\beta c_Z s_Z R_{1\alpha} 
+\frac{s_\beta}{4}\left[2\sqrt{3}(c_Z^2-s_Z^2)c_{331} +2(3c_{331}^2-1)c_Z^{} s_Z^{}\right]R_{2\alpha}\notag\\
&\hspace{2cm}+\frac{u}{4v}\left[-2\sqrt{3}(c_Z^2-s_Z^2)c_{331} +2(3c_{331}^2-1) c_Zs_Z\right]R_{3\alpha}
 \Big\}H_\alpha Z_\mu Z^{\prime\mu}\notag\\
&+ \frac{g^2v}{4}\sum_{\alpha=1-3}
\left(s_\beta R_{2\alpha} +\frac{u}{v} R_{3\alpha}\right)H_\alpha (Y_{1\mu} Y_1^\mu+Y_{2\mu} Y_2^\mu )\notag\\
&+\Big[\frac{g^2v}{6c_{331}}c_\beta s_\beta\left(\sqrt{3}c_Z^{} -3 s_Z^{}c_{331}\right)H^+W_\mu^-Z^\mu
-\frac{g^2v}{6c_{331}}c_\beta s_\beta\left(\sqrt{3}s_Z^{} + 3 c_Z^{}c_{331}\right)H^+W_\mu^-Z^{\prime \mu} \notag\\
&\quad\quad+\frac{g^2}{2}vc_\beta s_\beta H^+W_\mu^{\prime-}(Y_1^\mu -iY_2^\mu)\Big]+\text{h.c.}, 
\end{align}
where $c_Z^{}=\cos\theta_Z$ and $s_Z^{}=\sin\theta_Z$. 
Notice here that there appears the $H^\pm W^\mp Z$ coupling, which vanishes in THDMs at the tree level~\cite{Grifold,f,b1,b2,Yagyu}. 
Therefore, to measure this vertex is useful to discriminate our model from THDMs. The feasibility study of this vertex has been discussed 
at the LHC~\cite{HWZ-LHC} and at the ILC~\cite{HWZ-ILC}.  
In our model, however, we find that 
the coefficient of this vertex is only proportional to $v^2/u^2$ (plus the order $v^4/u^4$ correction) 
after taking the series expansion of the mixing angle $\theta_Z^{}$ under $v^2/u^2\ll 1$, so that 
the magnitude of this vertex is negligibly small.

\section{Higgs boson couplings to fermions}

The dimensionful $3\times 3$ couplings $\Gamma_q^\phi$ and $\Gamma_q^{H^\pm}$ given in Eq.~(\ref{yint}) are expressed as 
\begin{subequations}
\begin{align}
\Gamma_d^{H_\alpha}    &= V_L^d\text{diag}\left(\frac{R_{2\alpha}}{s_\beta},\frac{R_{2\alpha}}{s_\beta},\frac{R_{1\alpha}}{c_\beta}\right)(V_L^d)^\dagger M_d^{\text{diag}}, \\
\Gamma_u^{H_\alpha}    &= V_L^u\text{diag}\left(\frac{R_{1\alpha}}{c_\beta},\frac{R_{1\alpha}}{c_\beta},\frac{R_{2\alpha}}{s_\beta}\right)(V_L^u)^\dagger M_u^{\text{diag}}, \\
\Gamma_d^{A} &= \frac{i}{\sqrt{1 + s_\beta^2t_\gamma^2 }}V_L^d\text{diag}\left(-\frac{1}{t_\beta}, -\frac{1}{t_\beta},  t_\beta \right)(V_L^d)^\dagger M_d^{\text{diag}}, \\
\Gamma_u^{A} &= \frac{i}{\sqrt{1 + s_\beta^2t_\gamma^2 }}V_L^u\text{diag}\left(-t_\beta, -t_\beta, \frac{1}{t_\beta} \right)(V_L^u)^\dagger M_u^{\text{diag}}, \\
\Gamma_d^{H^\pm} &= V_L^u\text{diag}\left(\frac{1}{t_\beta}, \frac{1}{t_\beta},  t_\beta \right)(V_L^d)^\dagger M_d^{\text{diag}}, \\
\Gamma_u^{H^\pm} &= V_L^d\text{diag}\left(t_\beta, t_\beta, \frac{1}{t_\beta} \right)(V_L^u)^\dagger M_u^{\text{diag}}. 
\end{align}\label{eqs}
\end{subequations}
In Eq.~(\ref{eqs}), $V_L^d$ and $V_L^u$ are the unitary matrices which transform the weak eigenbasis of the left-handed quarks into 
the their mass eigenstates: $q_L^{} \to (V_L^q)^\dagger \, q_L^{}$ ($q=d,u$). 
$M_{d}^{\text{diag}}$ and $M_{u}^{\text{diag}}$ are the diagonalized mass matrices for the SM down- and up-type quarks, respectively. 
Notice here that in the above expressions, if the diag$(x,y,z)$ part is proportional to the $3\times 3$ identity matrix, 
we then obtain the same form of the Yukawa interaction as that in a $Z_2$ symmetric version of THDMs (see, e.g., \cite{AKKY}), where
the $V_L^q$ dependence disappears in the neutral Higgs boson couplings, and 
the CKM matrix $V_{\text{CKM}}\equiv V_L^u (V_L^d)^\dagger$ appears in the charged Higgs boson couplings. 
Consequently, the flavour violating quark Yukawa couplings to neutral Higgs boson do not appear at the tree level in the THDMs. 
However, this is not the case in our model, because at least the diag$(x,y,z)$ part for $A$ is not proportional to the identity matrix. 
As a result, the flavour violating couplings to the neutral Higgs bosons inevitably appear at the tree level, 
which is one of the most important consequences of the structure of our Yukawa interaction. 

\section{Decay rates of the Higgs bosons}

We present the analytic expressions for the decay rates of the extra Higgs bosons which are used to calculate the decay branching fractions 
as shown in Sec.~IV-B. 

The decay rates for the neutral Higgs bosons $\phi=A,H,h$ with a fermion pair in the final state are given as
\begin{align}
\Gamma(\phi \to q_i\bar{q}_j) &= N_c\frac{m_\phi^{}}{32\pi v^2}\Big\{
(1-x_{q_i}^2-x_{q_j}^2)\left(\left|(\Gamma_{q}^\phi)_{ij} + (\Gamma_{q}^\phi)_{ji}^*\right|^2+ \left|(\Gamma_q^\phi)_{ij} - (\Gamma_q^\phi)_{ji}^*\right|^2   \right)\notag\\
&-2x_{q_i}x_{q_j}\left(\left|(\Gamma_{q}^\phi)_{ij} +(\Gamma_q^\phi)_{ji}^*\right|^2 - \left|(\Gamma_q^\phi)_{ij} - (\Gamma_{q}^\phi)_{ji}^*\right|^2   \right) \Big\}
\lambda^{1/2}(x_{q_i}^2,x_{q_j}^2),  \label{qq}\\
\Gamma(\phi \to q\bar{q})&= N_c\frac{m_\phi^{}}{8\pi v^2}\Big\{
 (1-2x_q^2)\left[\text{Re}(\Gamma_{q}^\phi)^2 + \text{Im}(\Gamma_{q}^\phi)^2\right]\notag\\
 &\hspace{1cm}-2x_q^2\left[\text{Re}(\Gamma_{q}^\phi)^2 - \text{Im}(\Gamma_{q}^\phi)^2\right]\Big\}\beta(x_q^2), \\
 \Gamma(\phi \to \ell^+ \ell^-) &= \frac{m_\phi^{}}{8\pi v^2}m_\ell^2\,t^2_\beta\, \beta^{\,p_\phi}(x_\ell^2), 
\end{align}
where the two body phase space function $\lambda(x,y)$ is given by $\lambda(x,y)= 1+x^2+y^2-2x-2y-2xy$, and $\beta(x)=\sqrt{\lambda(x,x)}=\sqrt{1-4x^2}$. 
In the above expressions, we also introduced $x_a^{}=m_{a}/m_\phi$, $p_\phi=3\,(1)$ for $\phi=H\,(A)$, and the color factor $N_c$. 
For the expression of $\phi \to q_i\bar{q}_j$ mode given in Eq.~(\ref{qq}), 
the flavour index must not be identical, i.e., $i\neq j$. 
If the mass of $H$ is larger than $2\times m_h \simeq 250$ GeV, the $H\to hh$ decay channel also opens, and its decay rate is given by 
\begin{align}
\Gamma(H \to hh) = \frac{1}{8\pi}\frac{|\lambda_{Hhh}|^2}{m_H^{}}\sqrt{1-\frac{4m_h^2}{m_H^2}}, 
\end{align}
where $\lambda_{Hhh}$ is the coefficient of the $Hhh$ vertex in the Lagrangian. 
In the limit of $v/u\to 0$, we have 
\begin{align}
\lambda_{Hhh} = -\frac{c_{\beta-\alpha}}{2v}\left[m_h^2+m_A^2-m_H^2 +3(m_H^2-m_A^2)\frac{s_{2\alpha}}{s_{2\beta}} \right]. \label{Hhh}
\end{align}
The decay rate of the one-loop induced $\phi \to gg$ mode is given by 
\begin{align}
\Gamma(\phi \to gg)&=\frac{\sqrt{2}G_F\alpha_s^2 m_{\phi}^3}{128\pi^3 }
\left[\left|\sum_i \frac{\text{Re}(\Gamma_q^\phi)_{ii}}{m_{q^i}}F_1^{\phi}(m_{q^i})\right|^2 + \left|\sum_i \frac{\text{Im}(\Gamma_q^\phi)_{ii}}{m_{q^i}}F_2^{\phi}(m_{q^i})\right|^2\right], 
\label{gg}
\end{align}
where the loop functions are given by 
\begin{align}
F_{1}^{\phi}(m)  & = -\frac{4m^2}{m_{\phi}^2}\left[2-m_{\phi}^2\left(1-\frac{4m^2}{m_{\phi}^2}\right)C_0(0,0,m_{\phi}^2,m,m,m)\right], \notag\\
F_{2}^A(m)  & =-4m^2C_0(0,0,m_A^2,m,m,m),  \label{loopfunc}
\end{align}
with $C_0$ being the Passarino-Veltman three point scalar function~\cite{PV}. 

Finally, the decay rates for the charged Higgs boson $H^\pm$ into a pair of fermion is given by 
\begin{align}
\Gamma(H^+ \to u_i\, \bar{d}_j) &= N_c\frac{m_{H^\pm}}{16\pi v^2}\Big\{
(1-y_{u_i}^2-y_{d_j}^2)\left(\left|(\Gamma_{d}^{H^\pm})_{ij} + (\Gamma_{u}^{H^\pm})_{ji}^*\right|^2+ \left|(\Gamma_d^{H^\pm})_{ij} - (\Gamma_u^{H^\pm})_{ji}^*\right|^2   \right)\notag\\
&\hspace{-1cm}-2y_{u_i}y_{d_j}\left(\left|(\Gamma_{d}^{H^\pm})_{ij} +(\Gamma_u^{H^\pm})_{ji}^*\right|^2 - \left|(\Gamma_d^{H^\pm})_{ij} - (\Gamma_{u}^{H^\pm})_{ji}^*\right|^2   \right) \Big\}
\lambda^{1/2}(y_{u_i}^2,y_{d_j}^2), \\
\Gamma(H^+ \to \ell^+\, \nu) &= \frac{m_{H^\pm}^{}}{8\pi v^2}m_\ell^2\,t_\beta^2\,(1-y_\ell^2)^2, 
\end{align}
where $y_a = m_a/m_{H^\pm}^{}$. 

\end{appendix}

\end{document}